\documentclass[journal]{IEEEtran}
\usepackage{cite}
\usepackage{amsmath,amssymb,amsfonts}
\usepackage{isomath}
\usepackage{upgreek}
\usepackage{hyperref}
\usepackage{graphicx}
\usepackage{textcomp}
\usepackage{amsthm}
\usepackage{array}
\usepackage{soul}
\setstcolor{red}
\usepackage{booktabs}
\usepackage{multirow}
\usepackage{algorithm} 
\usepackage{algpseudocode}
\usepackage[font=small,
    labelfont=bf,
    format=plain]{caption}
\usepackage{xcolor}

\newcommand{\x}{\mathbf{x}}
\newcommand{\y}{\mathbf{y}}
\newcommand{\z}{\mathbf{z}}
\newcommand{\f}{\mathbf{f}}
\newcommand{\g}{\mathbf{g}}
\newcommand{\R}{\mathbb{R}}

\newcommand\numberthis{\addtocounter{equation}{1}\tag{\theequation}}
\newcommand{\vectheta}{\boldsymbol{\uptheta}}

\renewcommand{\vec}[1]{\ensuremath{\mathbf{#1}}}

\providecommand{\norm}[1]{\ensuremath{\left\lVert#1\right\rVert}}

\def\[{\begin{equation}}
\def\]{\end{equation}}
\DeclareMathOperator*{\argmin}{arg\,min}

\theoremstyle{definition}
\newtheorem{definition}{Definition}[section]

\theoremstyle{notation}

\theoremstyle{remark}

\theoremstyle{theorem}

\def\equationautorefname~#1\null{Eq.~(#1)\null}

\ifCLASSINFOpdf
\else
\fi
\begin{document}
%
\title{Compressible Latent-Space Invertible Networks for Generative Model-Constrained \\ Image Reconstruction}
%
%
%

\author{Varun A. Kelkar, Sayantan Bhadra, and Mark A. Anastasio, \IEEEmembership{Senior Member, IEEE}
\thanks{This work was supported in part by NIH Awards EB020604, EB023045, NS102213, EB028652, and NSF Award DMS1614305.}
\thanks{Varun A. Kelkar is with the Department of Electrical and Computer Engineering, University of Illinois Urbana-Champaign, Urbana, IL 61801 USA (e-mail: vak2@illinois.edu). }
\thanks{Sayantan Bhadra is with the Department of Computer Science and Engineering, Washington University in Saint Louis, Saint Louis, MO USA (e-mail: sayantanbhadra@wustl.edu).}
\thanks{Mark A. Anastasio is with the Department of Bioengineering, University of Illinois Urbana-Champaign, Urbana, IL 61801 USA (e-mail: maa@illinois.edu).}
\thanks{This paper has supplementary downloadable material available at \url{https://ieeexplore.ieee.org/}. The material includes additional results and figures. This material is 4.5 MB in size.}}

\markboth{IEEE Transactions on Computational Imaging,~Vol.~{}, No.~{}, December~2020}%
{Shell \MakeLowercase{\textit{et al.}}: Bare Demo of IEEEtran.cls for IEEE Journals}

\onecolumn
{\large IEEE Copyright Notice:}\\

$\copyright$ 2020 IEEE. Personal use of this material is permitted. Permission from IEEE must be obtained for all other uses, in any current or future media, including reprinting/republishing this material for advertising or promotional purposes, creating new collective works, for resale or redistribution to servers or lists, or reuse of any copyrighted component of this work in other
works.\\

This work has been submitted to the IEEE for possible publication. Copyright may be transferred without notice, after which
this version may no longer be accessible.

\twocolumn

\maketitle

\begin{abstract}
There remains an important need for the development of image reconstruction methods that can  produce diagnostically useful images from undersampled measurements. In magnetic resonance imaging (MRI), for example, such methods can facilitate reductions in data-acquisition times. Deep learning-based methods hold potential for learning object priors or constraints that can serve to mitigate the effects of data-incompleteness on image reconstruction. One line of emerging research involves formulating an optimization-based reconstruction method in the latent space of a generative deep neural network.  However, when generative adversarial networks (GANs) are employed, such methods can result in image reconstruction errors if the sought-after solution does not reside within the range of the GAN. To circumvent this problem, in this work, a framework for reconstructing images from incomplete measurements is proposed that is formulated in the latent space of  invertible neural network-based generative models. A novel regularization strategy is introduced that takes advantage of the multiscale architecture of certain invertible neural networks, which can  result in improved reconstruction performance over classical methods in terms of traditional metrics. The proposed method is investigated for reconstructing images from undersampled MRI data. The method is shown to  achieve comparable performance to a state-of-the-art generative model-based reconstruction method while benefiting from a deterministic reconstruction procedure and easier control over regularization parameters. 
\end{abstract}


%
\IEEEpeerreviewmaketitle

\section{Introduction}\label{sec:intro}

Modern imaging systems are typically computed in nature
and utilize a reconstruction method to estimate  an
image from a collection of measurements. 
In magnetic resonance imaging (MRI)  and other medical imaging modalities,
there are compelling reasons for reducing data-acquisition times.
In certain modalities, one way to achieve this is to simply reduce the number
of measurements acquired.
This strategy for accelerating data-acquisitions is relevant to MRI,
where data-acquisition times are proportional to the number of measured k-space samples.
 When the acquired  measurements are
insufficient to uniquely specify the sought-after object, i.e., the measurements are \emph{incomplete},
prior information about the object generally needs to be imposed in
 the form of regularization in order to recover images that possess potential
utility.

The concept of \textit{sparsity} has been widely exploited to develop effective
regularization strategies that can mitigate the effects of measurement-incompleteness in inverse
problems such as image reconstruction \cite{donoho_sparsity, donoho_sparsity2, candes_romberg_tao, pan}.
Modern sparse image reconstruction methods exploit the fact that many objects of
interest can typically be described by use of sparse representations and
have proven to be highly effective at estimating images from under-sampled
measurement data in MRI and other modalities  \cite{candes_review, sparsemri, mri_lustig, single_pixel}.
Sparse reconstruction  methods are commonly formulated as penalized least squares
estimators, where the penalty is specified as an $\ell_1$-norm that promotes solutions
that are sparse in a specified transform domain.
Such reconstruction approaches are prescribed by compressive sensing theories \cite{candes_robust, candes_romberg_tao, miki_elad_book}
and have enabled design of innovative  measurement strategies \cite{single_pixel, mri_lustig, yonina_sampler2}.


Instead of using hand-crafted penalties (i.e., object priors)
such as the $\ell_1$-norm or total variation semi-norm \cite{tvnorm}, there  has been considerable research aimed at
 learning object priors from a dataset of representative objects.
 Some of these techniques such as \textit{dictionary learning} and \textit{transform learning}, involve learning a dictionary that maps the images of interest to sparse vectors \cite{dictl_mri, tl_mri}. A more detailed review of sparsity and data-driven methods for image reconstruction can be found in reference \cite{sravi_review}. 
More recently, there have emerged numerous deep learning-based approaches for image reconstruction that
also seek to capture and exploit information regarding the sought-after object  in order to mitigate measurement-incompleteness or noise \cite{anastasio_dl, unet_recon, dip, bora, admmnet, manifold, vnn_mri, deepcascade}.

Deep generative models, such as generative adversarial networks (GANs), have shown
 great promise in learning  distributions of objects \cite{goodfellow, progan}.
 An object distribution represented by a generative model
 can be employed as an object prior in an image reconstruction approach, which can potentially
outperform traditional sparsity-based priors. For example, Bora \textit{et al.} proposed an approach 
in which the solution of a least squares image reconstruction problem
 is constrained
 to reside within the range of a generative model \cite{bora}.
This approach promotes solutions that are consistent with the measured data,
 with theoretical guarantees on the reconstruction error obtained \cite{bora}.
 It was also demonstrated empirically that in the severe undersampling regime, this method could outperform traditional
 sparsity-based reconstruction methods in terms of mean-squared error.
 Although this method can perform well when the measured data are produced from an object contained in the
range of the generator, in practice this condition can easily be violated.
Due to the GAN architecture, dataset size and variability, and other reasons outlined in \cite{aliahmed}, a high-dimensional object may not exactly reside on the low-dimensional manifold that
 is the range of the state-of-the-art GANs.
 This often leads to \textit{representation error}  and 
can result in reconstructed images that look realistic but contain false features.
 This phenomenon is highly undesirable in medical imaging applications.

Several approaches have been proposed to mitigate representation error in generative model-constrained
image reconstruction approaches.
One such approach, the \textit{SparseGEN} framework \cite{sparsegen}, accounts for sparse deviations of the true image from the range of the generative model, and achieves theoretical guarantees for signals that are only sparsely outside the range of a generative model. However, in practice, it might be difficult to find a linear mapping that sparsifies the difference of two realistic images. Another approach, known as the deep image prior (DIP) involves starting out with an untrained neural network and learning the parameters during reconstruction \cite{dip, csdip}. This method shows impressive performance, due to the fact that the structure of the convolutional neural network layers itself acts as a regularization. However, it has been shown that this approach eventually overfits the measurement noise, and early stopping is needed \cite{csdip}. A similar approach, known as image-adaptive GAN based reconstruction, starts out with a pretrained network similar to \cite{bora}, and then adapts the parameters of the GAN along with optimizing over the latent space vector \cite{iagan, iagan_sayantan, iagan_nyu}. These approaches involve optimizing over potentially a large number of parameters, depending upon the complexity of the GAN architecture.

A recent and promising approach to mitigating representation error in generative model-constrained
image reconstruction is to employ invertible neural networks (INNs) \cite{aliahmed}.
In INN-based generative models, referred to here as invertible generative models, the latent space and range have the same dimension
 and all possible images reside within the range.
 There also exists a unique latent space representation for every image.
 Hence, invertible generative models have theoretically zero representation error.
While the ability of invertible generative models to eliminate representation error
 is desirable,  an undesirable consequence of this is that they
 can also describe features that are not contained within the distribution of objects under
consideration. 
In this sense, the flexibility they provide in representing objects comes at the cost
of weakening the strength of the prior information employed to constrain the solution of the inverse problem.
This can result in reconstruction methods that produce object
estimates that are noisy or contain hallucinations \cite{aliahmed}.

In this work, novel regularization strategies are proposed for
 reconstruction methods that are constrained by use of
 invertible generative models.
Specifically, to address the limitations described above,
 the proposed regularization strategies are based upon the multiscale architecture of certain INNs.
 It is demonstrated that INNs with a multiscale architecture have a compressible
 latent space that can be exploited to effectively regularize the constrained image reconstruction
problem, resulting in images
 comparable to state-of-the-art image adaptive GAN based approaches 
while  benefiting
 from a deterministic reconstruction procedure and easier control over regularization parameters. 
While the proposed method is applicable to a variety of linear inverse problems, in this work,
 it is systematically investigated by means of stylized undersampled MRI experiments and compared to existing sparsity-based and generative model-based approaches.
This includes an investigation of 
 \textit{in-distribution} cases, where a test image belongs to the same probability distribution as the training data, and an \textit{out-of-distribution} case, where the image belongs to a distribution different from the training data. Finally, the proposed method is validated by use of a bias-variance analysis and 
other standard evaluation metrics such as mean-squared error and structural similarity \cite{ssim}. 

The remainder of the article is organized as follows. First, in \autoref{sec:bkd}, the considered problem is formulated and a description of compressed sensing under sparsity priors and using generative models is reviewed. In the same section, a brief introduction to invertible neural network architectures is provided. In \autoref{sec:comp_ls}, a description of how the latent space of certain INN architectures is compressible is given. A new reconstruction method that exploits the compressibility of the latent space for regularization is described in \autoref{sec:reg_compressibility}. The design of the numerical studies based on stylized MRI experiments is described in \autoref{sec:num_studies}, with the results given in \autoref{sec:results}. Finally, a discussion and conclusion is provided in \autoref{sec:discussion}.
\vspace{-0.2mm}
\section{Background}\label{sec:bkd}

Many digital imaging systems, including MRI, are well-approximated by a linear imaging model described as \cite{barrett}
\begin{align}\label{eqn:img_system}
    \g = H\f + \mathbf{n},
\end{align}
where $\f \in \mathbb{E}^n$ corresponds to the discretized approximation of the object to-be-imaged, $\g \in \mathbb{E}^m$ corresponds to the measurements taken, $H \in \mathbb{E}^{m\times n}$ corresponds to the linear discrete-to-discrete operator that approximately describes the imaging system, and $\mathbf{n} \in \mathbb{E}^m$ is the measurement noise. Here, the symbol $\mathbb{E}$ is used to denote a Euclidean space, specifically $\R$ or $\mathbb{C}$. In this work, the problem of estimating $\f$ from incomplete measurements $\g$ is considered; namely, $H$ is assumed to be rank deficient. 

\subsection{Recovering sparse objects from underdetermined systems}
\textit{Compressed sensing} (CS) has emerged as a popular framework for recovering signals from an underdetermined linear system of equations. In compressed sensing, the prior information about the structure of the object $\f$ is imposed through sparsity in some domain. More specifically, if $H$ satisfies the \textit{restricted isometry property} (RIP) over the set of all $2k$-sparse matrices, then the recovery of any $k$ sparse vectors can be guaranteed \cite{candes_romberg_tao, candes_review}. Several matrices, such as random Gaussian sensing matrices of appropriate column length and independent and identically distributed (i.i.d.) elements, as well as random Fourier sampling matrices relevant for compressive MRI satisfy the RIP \cite{candes_romberg_tao}. Intuitively, this means that two objects that are sparse in some domain, give rise to measurements that are not close. Hence, if an object is sparse, under certain conditions, it can be recovered uniquely from noiseless measurements \cite{candes_romberg_tao}.

\begin{definition}[Restricted Isometry]
Let $S_{k}$ be the set of all $k$-sparse vectors in $\R^n$. The \textit{restricted isometry constant} is the smallest constant $\delta_k \in (0,1)$ that satisfies
\begin{align}
    (1-\delta_k) \norm{\f}_2^2 \leq \norm{H\f}_2^2 \leq (1+\delta_k)\norm{\f}_2^2,
\end{align}
for all $\f \in S_k$. $H$ is said to satisfy the RIP over $S_k$ if $\delta_k$ is not too close to 1 in a prescribed sense \cite{candes_romberg_tao}. 
\end{definition}

\subsection{Recovering objects using generative priors}
A data-driven framework for compressed sensing has been developed \cite{bora}, where instead of sparsity in some domain, the prior information about the object is expressed in terms of a generative model.

Let $G : \R^k \rightarrow \R^n$ be a generative model, typically a deep neural network, parametrized by a vector $\vectheta$. The parameters $\vectheta$ of the generative model are estimated by training the model on a dataset of images, such that if a $\z \in \R^k$ is sampled from a simple tractable distribution, such as $\mathcal{N}(0, I)$, then $G(\z)$ is approximately a sample from the distribution of images that make up the dataset. Here, $\z$ is also called the latent representation of $G(\z)$. Since many image data distributions are approximately low dimensional, an architecture with $k \ll n$ can work well for the purpose of image generation. State-of-the-art generative performance is achieved by progressively growing generative adversarial networks (ProGAN) \cite{progan} and its variants, such as StyleGAN \cite{stylegan, stylegan2}.

Similar to the RIP, Bora \textit{et. al} in \cite{bora} introduced the  \textit{set-restricted eigenvalue condition} (S-REC) in the context of compressed sensing using generative models (CSGM).
\begin{definition}[Set-restricted eigenvalue condition]\label{def:srec}
Let $S \subseteq \R^n$. For some constants $\gamma > 0$ and $\delta \geq 0$, a matrix $H \in \R^{m\times n}$ satisfies the set-restricted eigenvalue condition $\text{S-REC}(S, \gamma, \delta)$ if for any $\f_1, \f_2 \in S$,
\begin{align}
    \norm{H(\f_1 - \f_2)}_2 \geq \gamma\norm{\f_1 - \f_2}_2 - \delta.
\end{align}
\end{definition}

It has been shown that specific sensing matrices, such as certain i.i.d. random Gaussian matrices, satisfy the S-REC \cite{bora}. Note that, similar to the RIP, the interpretation of Definition \ref{def:srec} is that
if $S$ is the range of a generative model, then any two objects $\f_1, \f_2$ in the range of the generative model that are sufficiently far apart in terms of the $\ell_2$ distance give rise to imaging measurements that are also far apart (up to an error of $\delta$), if the sensing matrix obeys a suitable S-REC. This is not the case for arbitrary vectors $\f_1, \f_2 \in \R^n$ that may have very different components in the null space of $H$ while giving rise to measurements that are close.

For a generative model $G$ with latent space dimensionality $k$ and Lipschitz constant $L$, Bora, \textit{et al.} \cite{bora} showed that $O(k\log(Lr/\delta))$ measurements suffice to stably recover those signals in $\mathcal{R}(G)$ whose latent representation has an $\ell_2$-norm of at most $r$. This recovery guarantee is applicable to a solution of the following optimization problem:
\begin{align*}\label{eqn:csgm_nonlagr}
    \hat{\z} &= \argmin_{{\z}, {\norm{\z}\leq r}} \norm{\g - HG(\z;\vectheta) }_2^2, \\
    \hat{\f} &\equiv G(\hat{\z}; \vectheta),\numberthis{}
\end{align*}
where $\g = H\tilde{\f} + \mathbf{n}$ is the measurement corresponding to the unknown true object $\tilde{\f}$.
Here, $\mathcal{R}(G)$ is the range of $G$, defined as 

\[\mathcal{R}(G) \equiv \{G(\z) ~ s.t. ~ \z \in \R^k\}\].

In \cite{bora}, this problem is reformulated in the Lagrangian form as:
\begin{align*}\label{eqn:csgm}
    \hat{\z} &= \argmin_{\z} \norm{\g - HG(\z;\vectheta) }_2^2 + \lambda\norm{\z}_2^2, \\
    \hat{\f} &\equiv G(\hat{\z}; \vectheta),\numberthis{}
\end{align*}
where $\lambda \in \R^+$ is a regularization parameter used to implicitly impose the constraint $\norm{z} \leq r$. While this problem is non-convex, it has been empirically observed that gradient descent-based algorithms can find critical points that have a sufficiently low value of the objective to yield a reconstruction with low error \cite{bora, iagan, aliahmed}.
For images that lie in the range of $G$, this gives a reconstruction for which the $\ell_2$ error is only limited by the magnitude of measurement noise and the error due to non-convergence of the gradient-descent type algorithm used to approximately solve \autoref{eqn:csgm}.

However, when $\tilde{\f} \not\in \mathcal{R}(G)$, the reconstructed estimate of $\tilde{\f}$ contains an additional error, known as the \textit{representation error} \cite{bora}. Here, the representation error is defined as
\begin{align*}
    \rho_G(\tilde{\f}) \equiv \min_{\z} \norm{G(\z) - \tilde{\f}}_2.
\end{align*}
In practice, $G$ has limited representational capacity and is trained on limited training data by optimizing a non-convex objective with gradient-based methods. Also, $\mathcal{R}(G)$ is only a $k$-dimensional manifold in $\R^n$. Hence, there is a significant representation error even for in-distribution images. This, coupled with the fact that the generative models such as the ProGAN produce highly realistic images, can result in plausible but wrong solutions to \autoref{eqn:csgm}.

One natural extension to the optimization problem formulated in \autoref{eqn:csgm} is to adapt $\mathcal{R}(G)$ based on the measured data. This can be achieved by jointly optimizing over the parameters $\vectheta$ of the generative model and the latent space vector $\z$:
\begin{align*}\label{eqn:iagan}
    \hat{\z}, \hat{\vectheta} &= \argmin_{\z, \vectheta} \norm{\g - HG(\z; \vectheta) }_2^2, \\
    \hat{\f} &\equiv G(\hat{\z}, \hat{\vectheta}).\numberthis{}
\end{align*}
This technique is known as image-adaptive GAN-based reconstruction (IAGAN) \cite{iagan}. Approximate solutions to \autoref{eqn:iagan} can be obtained using a standard gradient-descent based algorithm. However, as shown in \cite{csdip}, the success of such an algorithm depends upon early stopping, and convergence results in overfitting the noisy measurements. Moreover, the method is typically initialized with a solution of \autoref{eqn:csgm} \cite{iagan, iagan_sayantan}. Equation (\ref{eqn:csgm}) may need to be solved several times with different random initializations to yield an initial estimate that gives competitive performance for \autoref{eqn:iagan}, in terms of mean squared error.

\subsection{Invertible generative models}
\begin{figure}[!t]
\centerline{\includegraphics[width=1.05\linewidth]{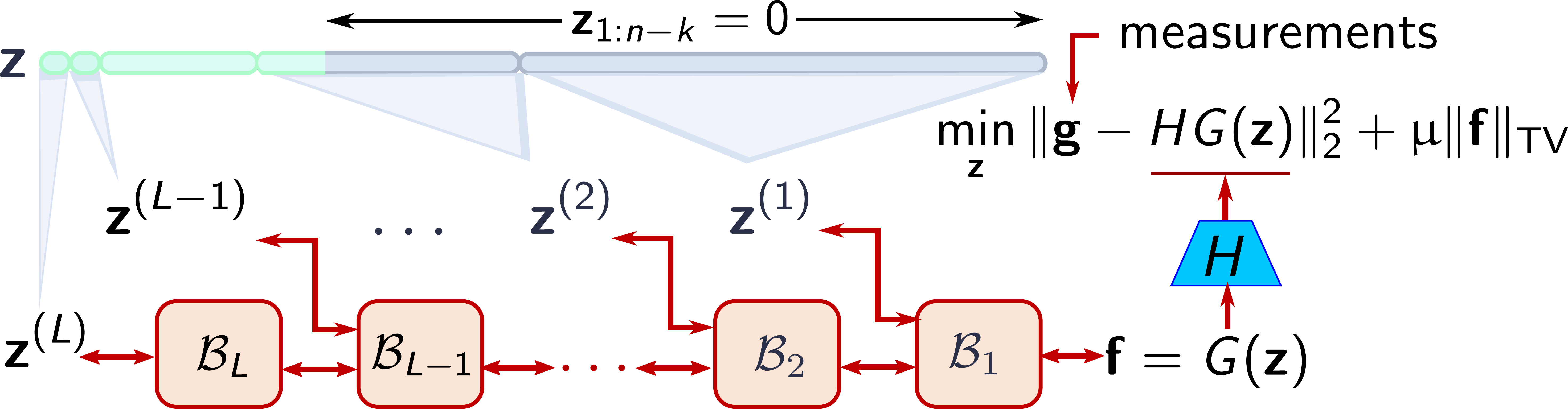}}
\caption{A schematic of our approach that exploits the multiscale structure of the INN. $\z^{(1)}, \z^{(2)}, \dots, \z^{(L)}$, are sections of the latent space that are introduced at different levels in the INN architecture, with $L$ being the number of levels. $\mathcal{B}_i$'s represent the invertible blocks that constitute the INN and $H$ is the forward model.}
\label{fig:multiscale}
\end{figure}
\begin{figure}[!t]
\centerline{\includegraphics[width=\linewidth]{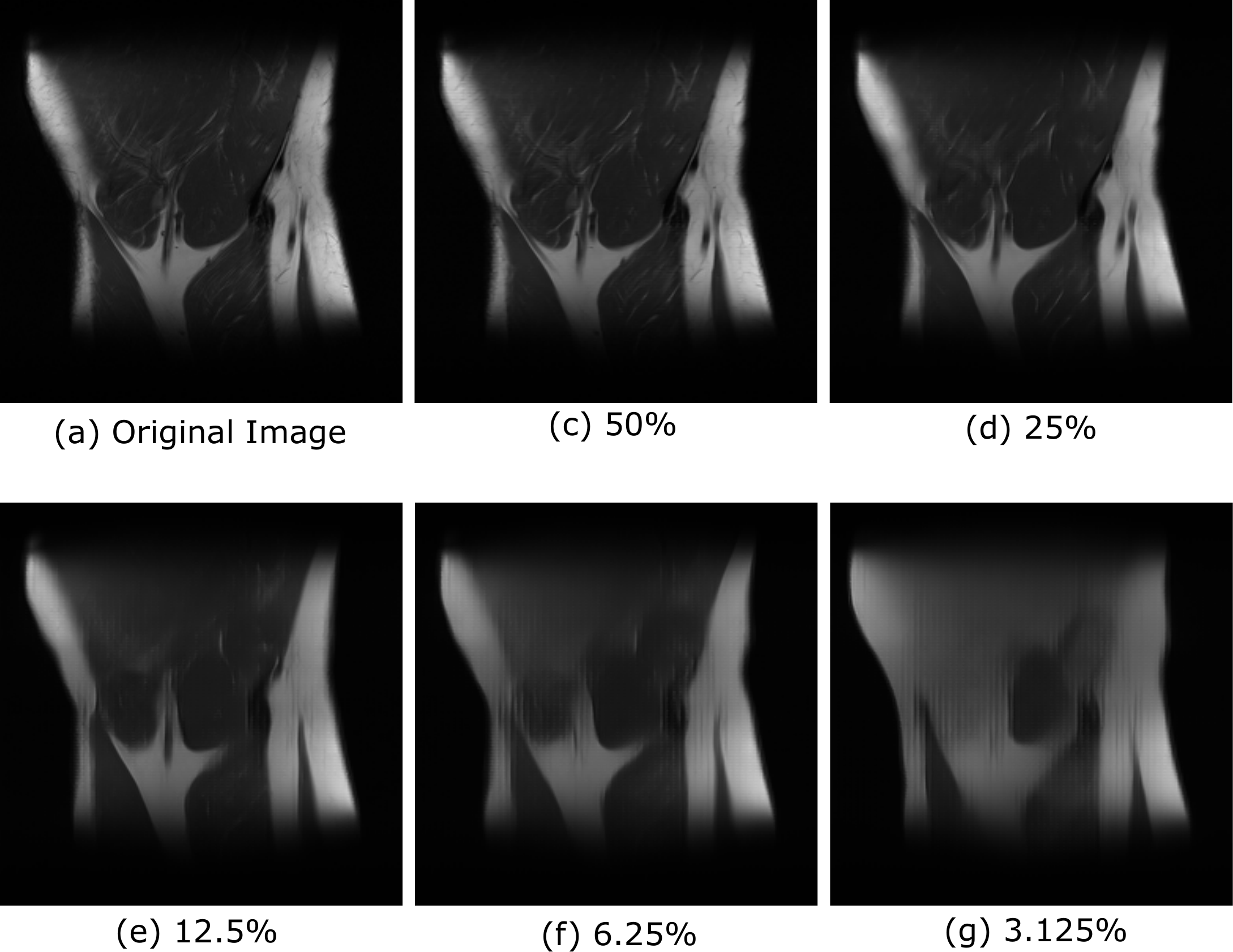}}
\caption{Low errors when $\z$ coefficients are truncated: a consequence of the compressible latent space. (a) A ground truth image, (b-f) images obtained by keeping a fraction of the $\z$ coefficients.}
\label{fig:truncation}
\end{figure}
Invertible neural networks (INN) are bijective mappings
\[G_{\rm{inn}} : \R^n \rightarrow \R^n,\]
constructed via neural networks, with the vector $\vectheta$ again denoting the parameters of the network \cite{normflow_review1, normflow_review2, dinh14, realnvp, glow, iresnet}. They can be trained as generative models on a dataset of images independently sampled from an image distribution $p_{\rm{\sf{f}}}$, such that a sample $\z \in \R^n$ from a simple tractable distribution $p_{\rm{\sf{z}}}$ produces a sample $\f = G_{\rm{inn}}(\z) \in \R^n$ from a distribution approximating $p_{\rm{\sf{f}}}$. Since the mapping is a bijection, every $\f$ has a unique latent-space representation $\z$. Moreover, the probability distributions of the input and the output of $G_{\rm{inn}}$ are related by \cite{normflow_review1, normflow_review2}
\begin{align}
    p_{\mathbf{\sf{f}}}(\f) |\det(\nabla_\z G_{\rm{inn}}(\z))| = p_{\mathbf{\sf{z}}}(\z),
\end{align}
or equivalently,
\begin{align}
    -\log p_{\mathbf{\sf{f}}}(\f) &= -\log p_{\mathbf{\sf{z}}}(\z) + \log |\det(\nabla_\z G_{\rm{inn}}(\z))|,
\end{align}
where $\f = G_{\rm{inn}}(\z)$ or, equivalently, $\z = G_{\rm{inn}}^{-1}(\f)$.

\begin{table}[!t]
\centering
\caption{Ensemble RMSE between the 500 test dataset images and images obtained by keeping a fraction of the $\z$ coefficients}
\label{tab:truncation}
\begin{tabular}{@{}cccccc@{}}
\toprule
\% $z_i$'s kept         & 50        & 25       & 12.5       & 6.25      & 3.125  \\
Mean RMSE         & 0.0090    & 0.0148   & 0.0244     & 0.0367    & 0.0518 \\ 
Std. dev. of RMSE    & 0.0024    & 0.0035   & 0.0057     & 0.0086    & 0.0135 \\ 
\bottomrule
\end{tabular}
\end{table}
Accordingly, an INN-based generative model can be trained by use of a log-likelihood based objective function: 
\begin{align*}
    \mathcal{L}(\mathcal{D}) &= -\frac{1}{D}\sum_{i=1}^{D} \log p_{\mathbf{\sf{f}}}(\f^{\{i\}})\\
    &= \frac{1}{D}\sum_{i=1}^{D}\log |\det(\nabla_\z G_{\rm{inn}}(\z^{\{i\}}))| - \frac{1}{D}\sum_{i=1}^{D} \log p_Z(\z^{\{i\}}),\numberthis{}\label{eqn:inn_training}
\end{align*}
where $\mathcal{D} = \{\f^{\{i\}}\}_{i=1}^D$ is the training dataset of size $D$.

For training scalable invertible networks via \autoref{eqn:inn_training}, the following conditions need to be satisfied:
(1) for an invertible layer that maps a vector $\x \in \R^n$ to a vector $\y \in \R^n$, computing $\x$ from $\y$ and computing $\y$ from $\x$ must have similar computational costs, and (2) the determinant of the Jacobian of the network is computationally tractable.
Several architectures satisfying the above constraints have been proposed \cite{dinh14, realnvp, glow, flowpp}. In many of these architectures, a key enabling factor in satisfying the above constraints is the affine coupling layer. If $\x$ and $\y$ are the input and output of a certain affine coupling layer, the transformation relating $\y$ to $\x$ is given by
\begin{align*}\label{eqn:affine_coupling}
    \y_{1:p} &= \x_{1:p},\\
    \y_{p+1:n} &= \x_{p+1:n} \odot \exp(s(\x_{1:p})) + t(\x_{1:p}),\numberthis{}
\end{align*}
where the notation $\x_{u:v}$ is used to denote the vector formed from components of $\x$ from the $u$-th index to the $v$-th index, and $\odot$ denotes the Hadamard product or the element-wise product of two vectors. Here, $s$ and $t$ are functions parametrized by neural networks, and need not be invertible. It can be verified that the above transformation is invertible \cite{realnvp}. Moreover, the determinant of the Jacobian of the above transformation is given by
\begin{align}
    \left|\det\left( \frac{\partial \y}{\partial \x} \right)\right| = \exp\left( \sum_{i=1}^{n-p} s(\x_{1:p})_i \right),
\end{align}
which is computationally inexpensive as compared to the usual $O(n^3)$ complexity needed to evaluate a general $n\times n$ Jacobian.

The use of INNs to impose a reconstruction constraint can be achieved by replacing the GAN-based generative model in \autoref{eqn:csgm} with an invertible network \cite{aliahmed}. However, in current practice, the state-of-the-art GANs generally model the probability distribution of the data better than the state-of-the-art invertible generative models, both perceptually, as well as in terms of the Fr\'echet Inception Distance (FID) scores between real and generated datasets \cite{fid}. Hence, a naive application of INNs in \autoref{eqn:csgm} can result in a noisy or distorted image estimate when dealing with high resolution images. Intuitively, \autoref{eqn:csgm} can be thought of as optimizing over the norm ball $\norm{\z}\leq r \in \R^{+}$ which, when high dimensional, contains undesirable solutions to \autoref{eqn:csgm}. Hence, a new way of regularizing the problem is needed in order to constrain the solution space while minimizing the representation error.

\section{Compressibility of INN Latent Space}\label{sec:comp_ls}
\begin{figure*}[!t]
\centerline{\includegraphics[width=\linewidth]{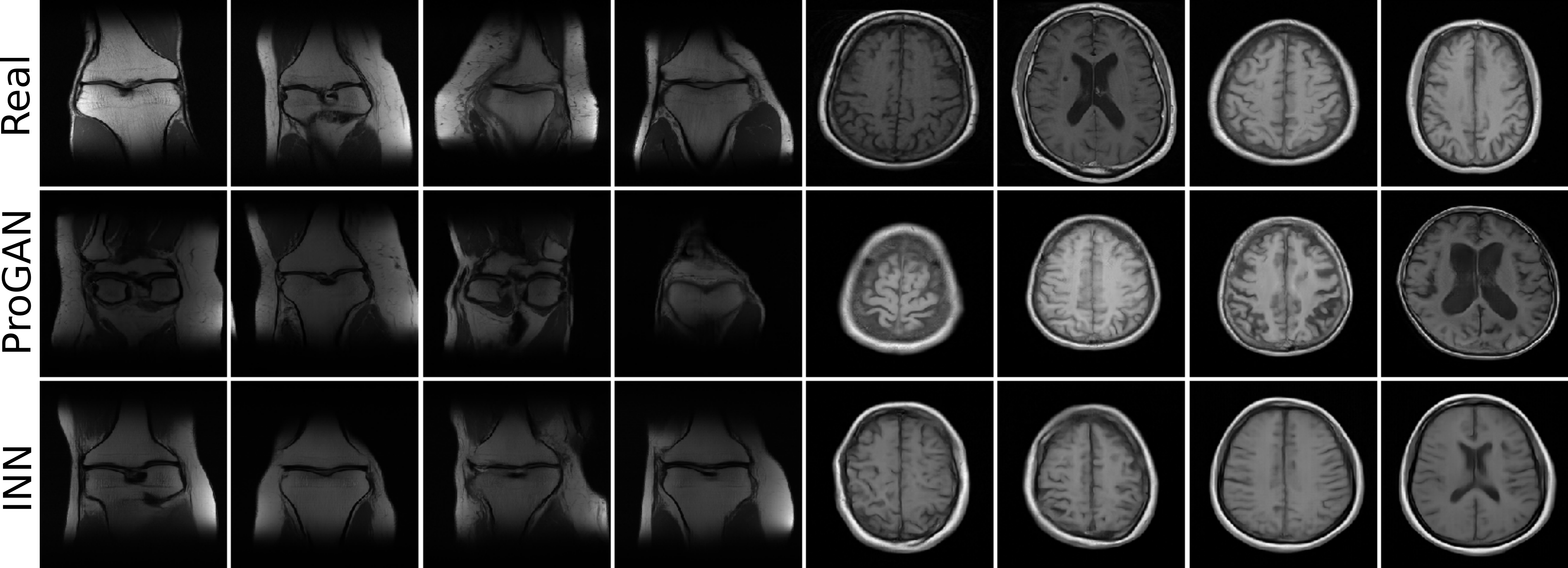}}
\caption{Generated samples from the progressive GAN and the INN, alongside the experimentally acquired ``real" images.}
\label{fig:samples}
\end{figure*}
\begin{figure*}[!t]
\centerline{\includegraphics[width=\linewidth]{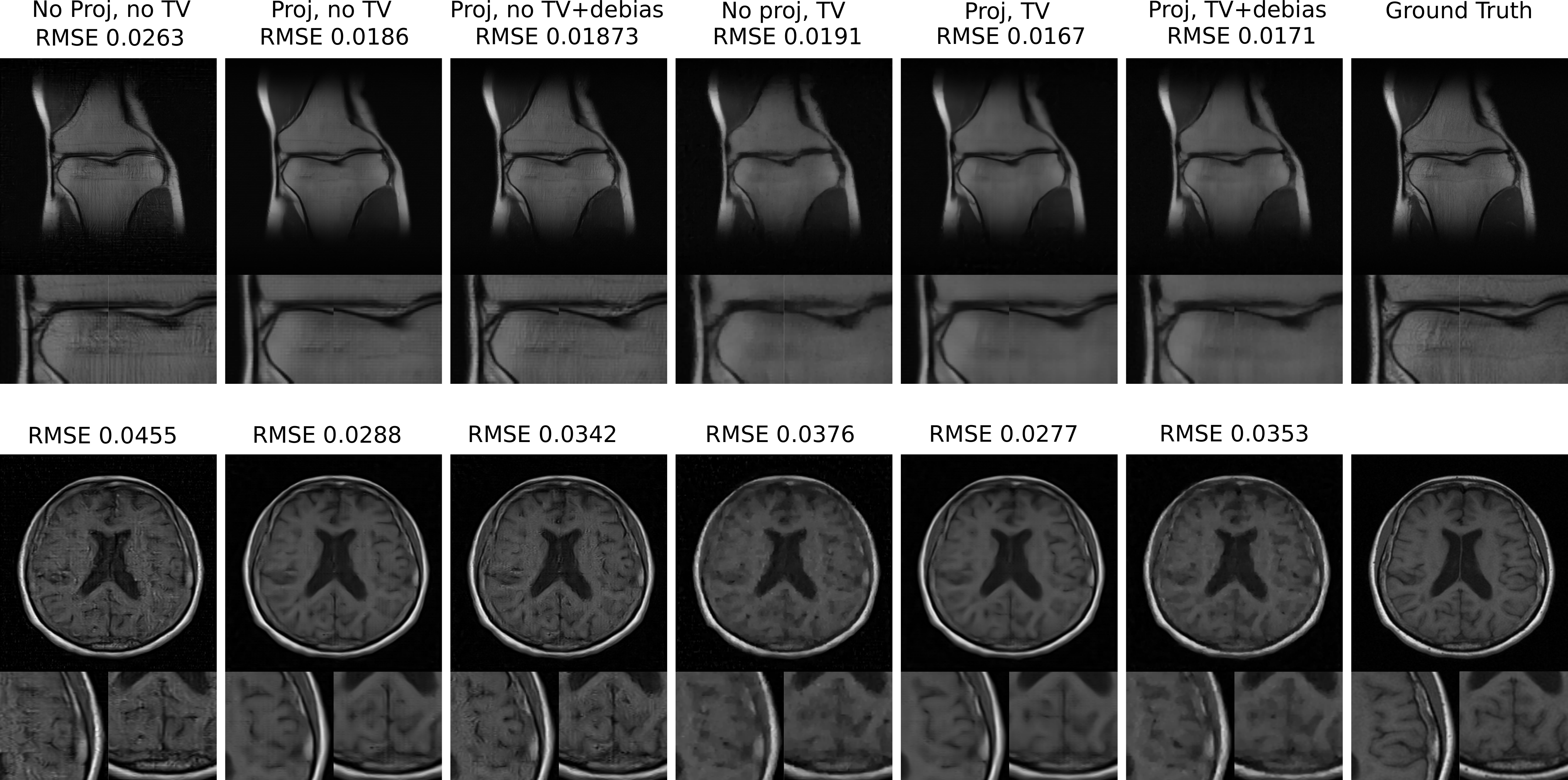}}
\caption{Comparison of images reconstructed by use of various regularization combinations in the INN reconstruction framework.}
\label{fig:proj_comparison}
\end{figure*}

\begin{figure}[!t]
\centerline{\includegraphics[width=\linewidth]{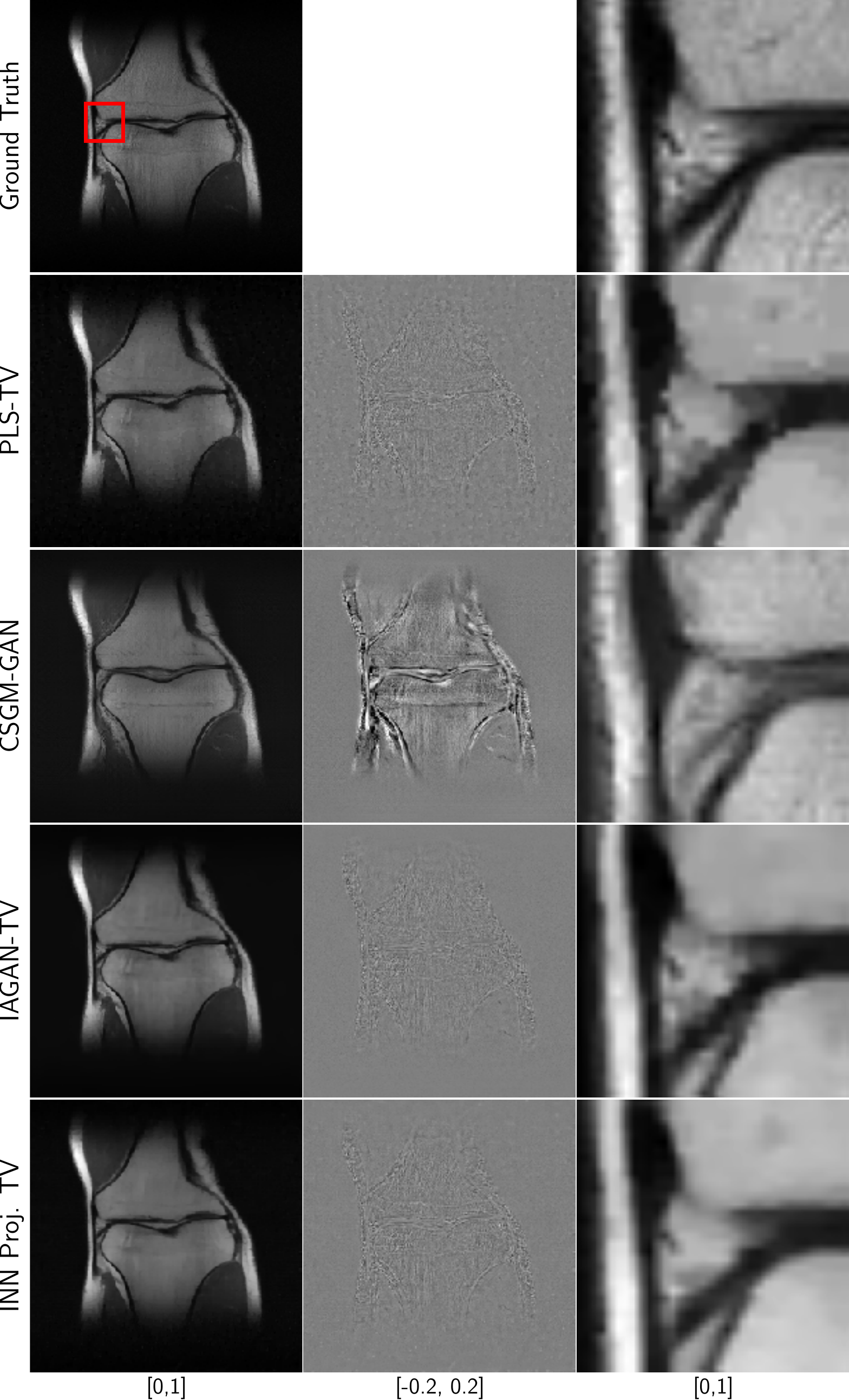}}
\caption{Ground truth, difference plots and reconstruction results for a coronal PD weighted knee image without fat suppression, with 8-fold undersampling and 20 dB measurement SNR.   The RMSE and SSIM values are displayed in \autoref{tab:mse_ssim}.}
\label{fig:knee8x}
\end{figure}
In order to reduce the computation and memory during training, Dinh, \textit{et al.} proposed a multiscale invertible architecture \cite{realnvp}. As shown in figure \autoref{fig:multiscale}, this results in sections of the latent space vector being introduced into the network at different points in the network, leading to a compressible latent space \cite{realnvp}.
This operation can be recursively described as \cite{realnvp}
\begin{align}
    \z &= ( \z^{(1)}, \z^{(2)}, \dots, \z^{(L)} ),\\
    \vec{h}^{(L-1)} &= \mathcal{B}_L(\z_L),\\
    \vec{h}^{(l)} &= \mathcal{B}_{l+1}(\vec{h}^{(l+1)}, \z^{(l+1)}), ~ l = 0,\dots, L-2,\\
    \f &= \vec{h}^{(0)},
\end{align}
where $\mathcal{B}_l$ represents the $l$-th invertible block, constructed out of affine coupling layers, and $L$ is the number of levels in the INN.
The schematic in \autoref{fig:multiscale} shows the compressible structure of the latent space. 
The effect of this compressibility was examined on an ensemble of 500 images from a test dataset of coronal knee images that was kept out of the INN training dataset.
This was done as follows. First, for an image $\f_{\rm{orig}}$ in the ensemble, the exact latent representation $\z_{\rm{orig}}$ was computed. This can be divided into multiple sections $\z^{(1)}, \z^{(2)}, \dots, \z^{(L)}$ based on the multilevel architecture of the INN. Next, all sections from $\z^{(1)} \dots \z^{(i)}$ were progressively set to zero such that only 50\%, 25\%, 12.5\%, 6.25\% and 3.125\% of the components of $\z$ remain non-zero. For these modified latent space vectors $\z_{50\%}$, $\z_{25\%}$, $\z_{12.5\%}$, $\z_{6.25\%}$ and $\z_{3.125\%}$, the corresponding images $\f_{P} = G_{\rm{inn}}(\z_{P}),~ P = 50\%,~ 25\%,~ 12.5\%,~ 6.25\% \text{ and } 3.125\%$ were computed, and the root mean square errors (RMSEs) $\norm{\f_P - \f_{\rm{orig}}}/\sqrt{n}$ with respect to the original image $\f_{\rm{orig}}$ were calculated. The error versus the percentage of $\z$ coefficients kept, averaged over the entire ensemble, was computed and is reported in \autoref{tab:truncation}. Figures \ref{fig:truncation}b-\ref{fig:truncation}f display the effect of the truncation on a single ground-truth image \autoref{fig:truncation}a.
Thus, in addition to being a generative prior, these results suggest that a multilevel INN architecture also serves to establish a compressible representation for in-distribution images, which is consistent with the findings in \cite{realnvp}. 
The performance of the INN in terms of compressibility was compared to that of the Haar wavelet transform via RMSE and SSIM, and is included in the supplementary section.

\section{Regularization through compressibility}\label{sec:reg_compressibility}
The fact that the latent space of the INN is compressible can be exploited to design a regularization strategy for inverse problems. Motivated by the compressible structure of the latent space, the recovery of objects $\f$ such that $\norm{G^{-1}(\f)}_1/n \leq r/n = 1 + o(1)$ will be examined. Let $S_k = \{ \f ~ s.t. ~ G^{-1}(\f)_{1:n-k} = 0 \}$, with $ k \in \mathbb{N}, ~ k \leq n$, and let $T_{\nu} = \{ \f ~ s.t. ~ \norm{\f}_{\rm{TV}} \leq \nu \}$, $\nu \in \R^{+}$. The relevant measurement model can be described as
\begin{align}\label{eqn:meas_model}
    \g = H\f, ~~ \text{where } \f \in S_k \cap T_{\nu}.
\end{align}

Note that the mapping in \autoref{eqn:meas_model} may not be injective. An approximate inverse of the above measurement model can be implicitly defined via a problem similar to the one in \autoref{eqn:csgm}, with an added constraint of restricting $\z$ to a $k$-dimensional subspace corresponding to the most important coefficients. A TV penalty on the image is also included. The considered optimization problem is stated as follows:
\begin{align*}\label{eqn:csgm_inn}
    \hat{\z} =& \argmin_{\z} \norm{\g - HG_{\rm{inn}}(\z) }_2^2 - \lambda\log p_{\sf{z}}(\z) \\
    &\qquad\qquad\qquad\qquad\qquad\qquad\quad + \mu\norm{G_{\rm{inn}}(\z)}_{\rm{TV}},\\
    \text{subject to} &\quad \z_{1:n-k} = 0,\\
    \hat{\f} \equiv&~ G_{\rm{inn}}(\hat{\z}),\numberthis{}
\end{align*}
where $k$ is used to restrict $\f$ to $S_k$, and $\lambda$ and $\mu$ are used to implicitly impose the constraints $\norm{\z}_1 \leq r$, and $\f \in T_{\nu}$ respectively. All of these are treated as regularization parameters which need to be tuned in order to achieve a suitable restriction. However, it was observed in preliminary studies that $\lambda$ is not critical to the reconstruction performance. In fact, the best results were achieved if $\lambda$ is set to 0, as shown in \autoref{tab:lamsweep}. This reduces the number of explicit regularization parameters to two - the dimensionality $k$ of the latent subspace and the TV penalty $\mu$. Note that similar to \autoref{eqn:csgm}, the objective function in \autoref{eqn:csgm_inn} is non-convex, and only approximate solutions are typically obtained when gradient-based methods are employed \cite{sidky_pan_cnn}. Moreover, a medical image may not lie in $S_k \cap T_{\nu}$; however, \autoref{eqn:csgm_inn} can potentially yield image estimates in $S_k \cap T_{\nu}$ that are close to the original object.

\begin{table}[!t]
\centering
\caption{RMSE values for reconstructions with varying $\lambda$}
\label{tab:lamsweep}
\begin{tabular}{@{}cccccc@{}}
\toprule
$\lambda$    & 0        & $10^{-4}$ & $3\times10^{-4}$ & $10^{-3}$ & $3\times10^{-3}$  \\
RMSE (\%)    & 1.342    & 1.349     & 1.353            & 1.390     & 1.535 \\ \bottomrule
\end{tabular}
\end{table}

Note that, due to the latent space projection, the range of the INN is restricted to $S_k$. Due to this, overfitting to noise can be avoided. However, this results in representation error, now defined as
\begin{align}
    \rho_{k}(\tilde{\f}) = \min_{\f \in S_k} \norm{\f - \tilde{\f}}_2.
\end{align}
Now, $\rho_k(\tilde{\f})$ is upper bounded by the truncation error,
\begin{align}
    \tau_k(\tilde{\f}) = \norm{\f_t - \tilde{\f}}_2,
\end{align}
where 
\begin{align}
\f_t = \argmin_{\f \in S_k} \norm{G_{\rm{inn}}^{-1}(\f) - G_{\rm{inn}}^{-1}(\tilde{\f})}.
\end{align}
According to the previously shown compressibility results, this error is expected to be minimal for in-distribution images, as compared to the representation error incurred in the approach described in \cite{bora}. 

Due to the restriction of the measurement operator on $S_k$, Lemma 4.1 in \cite{bora} implies that a random measurement matrix with $m = O(k/\alpha^2\log(Lr/\delta))$ rows and i.i.d. elements drawn from $\mathcal{N}(0, 1/m)$  satisfies the S-REC$(S_k, 1-\alpha, \delta)$ with high probability, where $L$ is now the Lipschitz constant of $G_{\rm{inn}}$, with $0<\alpha<1$ and $\delta > 0$. A large $L$ and the requirement of uniform recovery implies that this bound on the number of measurements needed is pessimistic for the type of images and measurement operators examined in this manuscript for both the invertible generative model, as well as the GAN. This is analogous to similar observations about RIP-based guarantees in general \cite{rip_absence, cs_theory_new}. However, in this work, this theoretical result is relevant because it provides intuitive understanding on how the number of measurements scale with respect to the dimensionality $k$ of the latent subspace that is used to restrict the domain of $H$.

A approximate solution of \autoref{eqn:csgm_inn} can be found by a regularized projected Adam technique, which includes iterative steps of the Adam algorithm \cite{adam} with a proximal step on $\z$ followed by a projection onto the convex subspace $\{\z ~|~ \z_{1:n-k} = 0  \}$ after each gradient step. The procedure for finding an approximate solution of \autoref{eqn:csgm_inn} is shown in \autoref{alg:projadam}.\\

\begin{algorithm}
	\caption{The proposed Projected Adam algorithm for finding an approximate solution to \autoref{eqn:csgm_inn}.}
    \label{alg:projadam}
	\begin{algorithmic}[1]
	\State \textbf{Given}: Measurements $\g$.
	\State Pick the regularization parameters $k$ and $\mu$. Fix $\lambda = 0$.
	\State Set $\mathbf{p}^{(k)} \in \R^n$ such that $\mathbf{p}^{(k)}_{1:n-k} = 0$ and $\mathbf{p}^{(k)}_{n-k+1:n} = 1$.
	\State Set the Adam optimizer parameters $(\alpha, \beta_1, \beta_2)$ from \cite{adam}. Default parameters recommended in Algorithm 1 of \cite{adam} are used.
	\State $\z_{\mathrm{init}} \leftarrow \mathbf{0}.$ (Initialize the latent space vector)\\
	$t \leftarrow 0.$ (Initialize the iteration number)
		\While {$\z_t$ not converged}
		    \State Perform an Adam iteration from Algorithm 1 in \cite{adam},
            $$\z_t \leftarrow \texttt{ADAM}_{\alpha, \beta_1, \beta_2}(\z_{t-1}; \mu),$$
            \State Perform projection onto the latent subspace:
            $$\z_t \leftarrow  \z_t \odot \mathbf{p}^{(k)} $$
            $$t \leftarrow t+1$$
		\EndWhile
	\end{algorithmic}
\end{algorithm}

\noindent\textit{Debiasing.}\label{sec:debiasing}
It was observed that the reconstruction accuracy of the proposed approach can be further improved by debiasing an approximate solution of \autoref{eqn:csgm_inn}. This can be achieved by removing the constraint of $\z$ lying in a $k$-dimensional subspace. Debiasing can be performed by finding an approximate solution to the following:
\begin{align*}\label{eqn:csgm_inn_debias}
    \hat{\z}_{\rm{deb}} &= \argmin_{\z} \norm{\g - HG_{\rm{inn}}(\z) }_2^2
    + \mu\norm{G_{\rm{inn}}(\z)}_{\rm{TV}} \\
    \hat{\f}_{\rm{deb}} &= G_{\rm{inn}}(\hat{\z}_{\rm{deb}}),\numberthis{}
\end{align*}
where the above problem is initialized with an approximate solution of \autoref{eqn:csgm_inn}.
Improvement via debiasing can be obtained through iterative minimization of \autoref{eqn:csgm_inn_debias} by use of early stopping. However, similar to other techniques that require early stopping \cite{iagan, dip}, use of an appropriate stopping criterion requires additional tuning parameters \cite{dip_automation}. Hence, debiasing with early stopping was not employed in the studies described below.

The proposed reconstruction method was evaluated as described below.

\section{Numerical Studies}\label{sec:num_studies}
\begin{figure}[!t]
\centerline{\includegraphics[width=\linewidth]{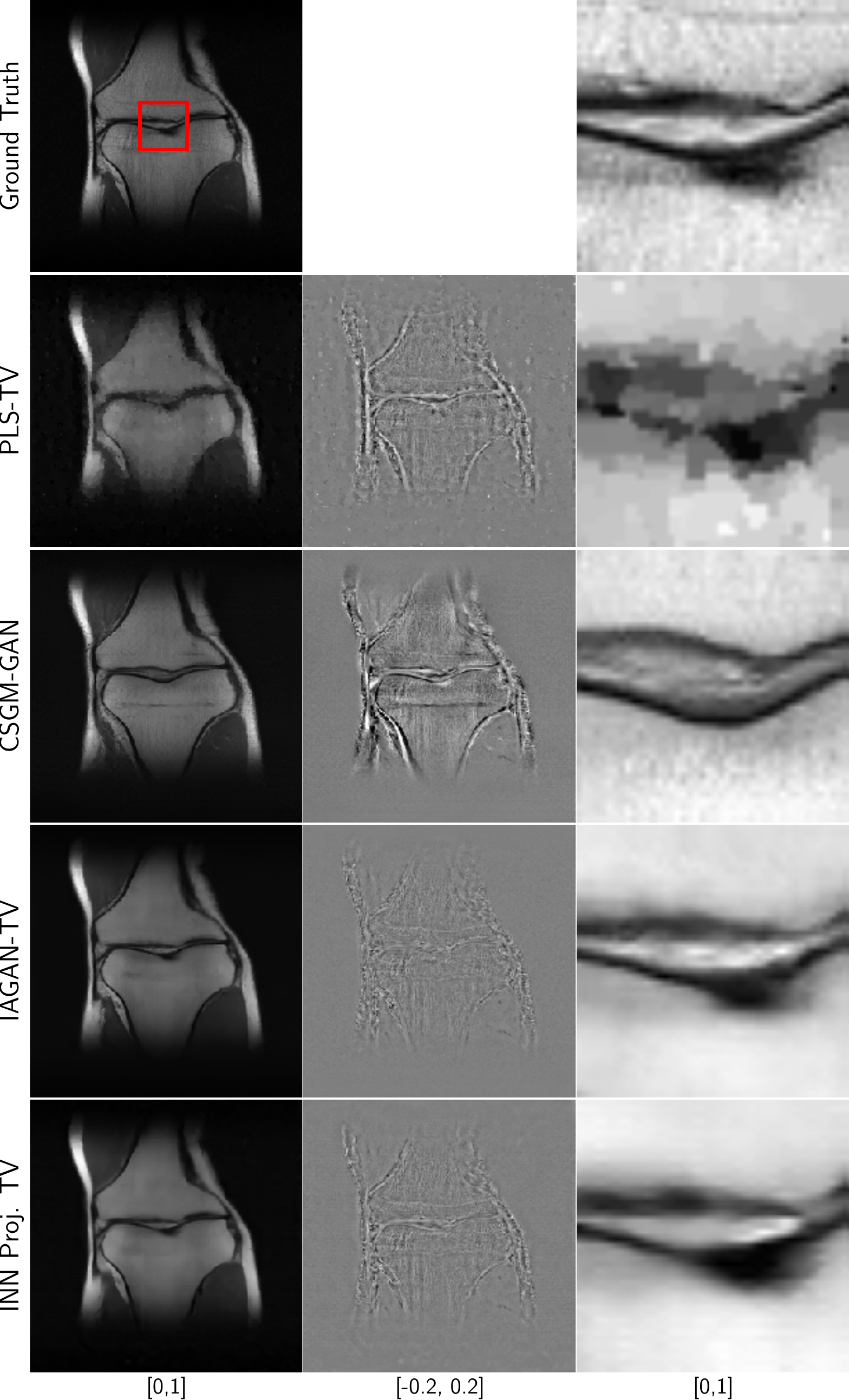}}
\caption{Ground truth, difference plots and reconstruction results for a coronal PD weighted knee image without fat suppression, with 20-fold undersampling and 20 dB measurement SNR. The RMSE and SSIM values are displayed in \autoref{tab:mse_ssim}.}
\label{fig:knee20x}
\end{figure}

\begin{figure}[!t]
\centerline{\includegraphics[width=\linewidth]{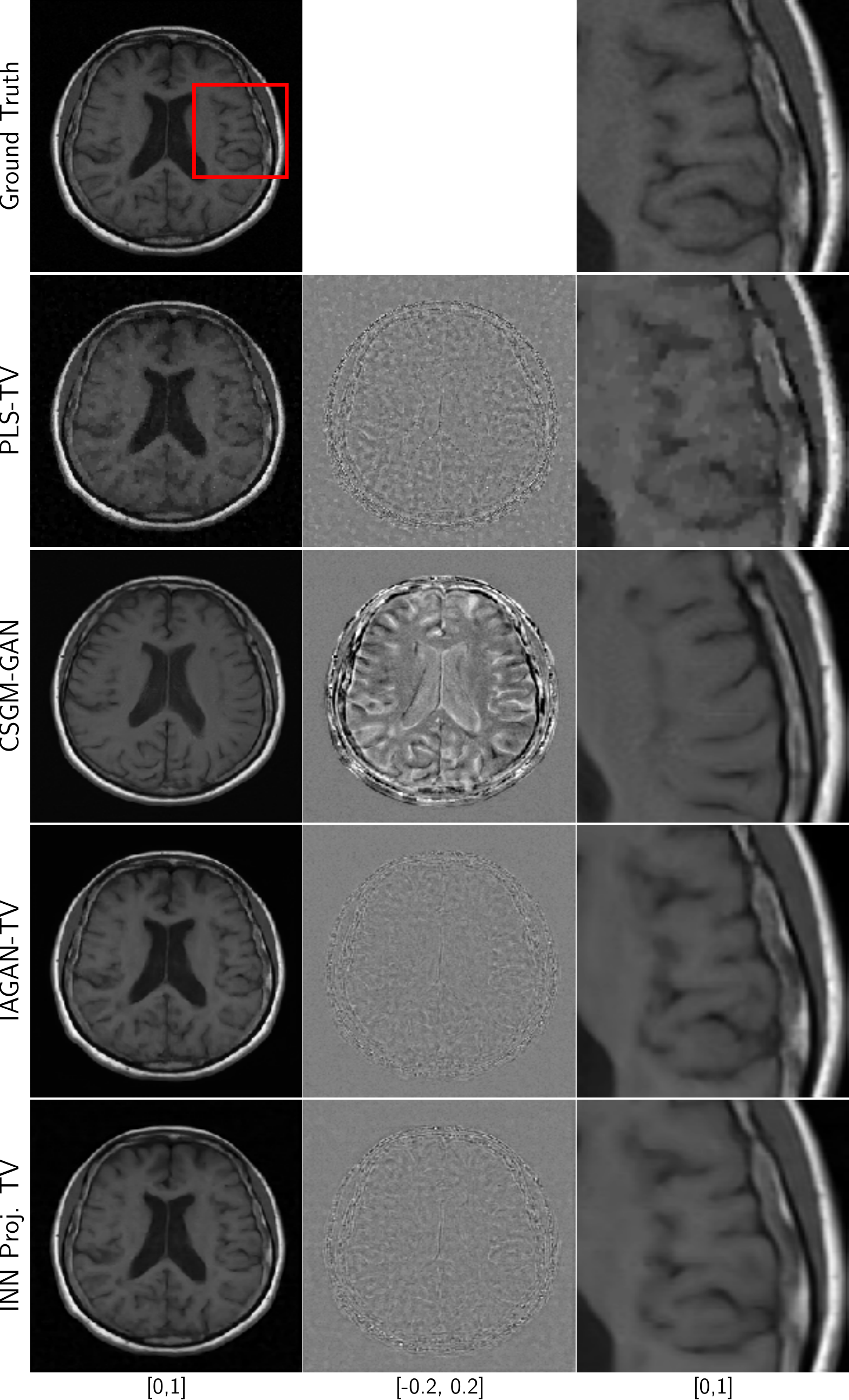}}
\caption{Ground truth, difference plots and reconstruction results for an axial T1 weighted brain image, with 8-fold undersampling and 20 dB measurement SNR. The RMSE and SSIM values are displayed in \autoref{tab:mse_ssim}.}
\label{fig:brain8x}
\end{figure}
Numerical studies were conducted to assess the effectiveness of the proposed method, especially in terms of recovering fine object features. The reduction in the appearance of realistic but false features and oversmoothing artifacts was studied. 
Our studies were divided into two parts - (1) reconstruction from stylized, simulated undersampled single-coil MRI measurements (henceforth referred to as the simulation study), and (2) reconstruction from emulated experimental undersampled single-coil MRI measurements (henceforth referred to as the emulated experimental study).
For the simulation study,
in-distribution images, i.e. the images that come from the same distribution as the training dataset, as well as out-of-distribution images were considered. The proposed method was compared to traditional sparsity-based, as well as recent GAN-based reconstruction methods. For the comparisons, traditional image quality metrics such as the root mean squared error defined as the discrete error norm $\|\hat{\f} - \tilde{\f}\|_2$, as well as structural similarity (SSIM) index \cite{ssim} were utilized. Where applicable, bias-variance tradeoff calculations were carried out to assess the robustness of the algorithms. 

\subsection{Datasets and sensing system}
\subsubsection{Simulation study}
The generative models were trained on single channel 2D MRI images of size 256 $\times$ 256. The following two datasets were employed for training, as well as evaluation in the case of in-distribution images:
\begin{itemize}
\item \textit{FastMRI knee dataset}:
15000 non-fat suppressed, proton density (PD) weighted coronal knee images from the NYU fastMRI Initiative database \cite{fastmri}. 
\item \textit{FastMRI brain dataset}:
12000 T1-weighted axial adult brain images from the NYU fastMRI Initiative database.
\end{itemize}

For evaluation of reconstruction performance on out-of-distribution images, images from a pediatric epilepsy resection MRI dataset containing anomalies \cite{paedomri, paedomri_dataset} were used, along with generative models trained on the FastMRI brain dataset.
Evaluating the robustness of a reconstruction method on out-of-distribution images is relevant because (i) in practice, test images may not exactly correspond to the training data distribution and a practitioner might be oblivious to these small differences, and (ii) it is of interest to examine the scenario of \textit{transfer compressed sensing}, where learned priors from one dataset are employed to recover images from a closely related but different test distribution, due to the unavailability of sufficient data to learn the priors (for example, data including rare anomalies.)

Simulated undersampled single-coil MR measurements were employed as a proxy for experimental MRI $k$-space measurements. Variable density Poisson disc sampling patterns shown in \autoref{fig:undersampling_masks} corresponding to R = 8 and R = 20 undersampling ratios were utilized, which retain low frequencies and randomly sample higher frequencies with a variable density \cite{bart, mri_lustig}. 

\subsubsection{Emulated experimental study}
Data for training the generative models were prepared in the following way. The fastMRI initiative database provides \textit{emulated single-coil $k$-space measurements}, each of which is a complex-valued linear combination of responses from multiple coils of raw multi-coil $k$-space data \cite{singlecoil_emulation}. These fully sampled $k$-space measurements were used to generate complex-valued images via the inverse fast Fourier transform (IFFT). They were divided into a training dataset for training the generative models, and a test dataset. The complex-valued images were converted to two-channel real images for training, and generative models with two-channel output were trained. Image reconstruction was performed directly from retrospectively undersampled emulated single-coil measurements, and the image estimates were compared with the reconstructions from the corresponding fully sampled k-space measurements in the test dataset. A Cartesian random undersampling mask with R = 4 was used for the retrospective undersampling.

For evaluating the reconstruction performance on the above-described image types,
a validation image and a test dataset was used for each of the image types.
These images were kept unseen during training.
The regularization parameters for all the reconstruction methods were tuned on the respective validation image for each image type, and the parameter setting showing the best RMSE performance was chosen. Images corresponding to the regularization parameter sweeps are shown in the supplementary section. The tuned parameters were used in the reconstruction of images from the unseen test datasets. Performance metrics and their statistical significance were reported on these test datasets for all the image types.

\subsection{Network architecture and training}
The employed INN architecture was adapted from Kingma and Dhariwal \cite{glow}. It consisted of 6 levels in the multilevel architecture, with this choice being empirically determined. The primary invertible layers used were affine coupling layers \cite{realnvp} with invertible 1x1 convolutions \cite{glow}. The functions $s$ and $t$ described in \autoref{eqn:affine_coupling} were parametrized by 3 layer convolutional neural networks, with SoftPlus activation functions as the nonlinearity between the convolutional layers \cite{softplus}. Also, similar to the official implementation by Kingma and Dhariwal \cite{glow_github}, the exponential function in the affine coupling layer was replaced with a sigmoid function, which stabilizes the training and makes the invertible transformation Lipschitz stable. In order to gain Lipschitz stability in the reverse direction, the output of the sigmoid function was rescaled to lie in the range $(c, 1]$ with $c$ being a positive, tunable parameter less than 1. Lastly, it was determined that using a standard i.i.d. Laplacian distribution as the latent space prior $p_{\mathsf{z}}(\z)$ improves the performance during image generation and reconstruction. Loosely speaking, the Laplacian prior, being more ``compressible" than the Gaussian prior, seems to help in learning a closer approximation to the near-low dimensional real data-distribution.
The INN was trained on a system with a 2x 20-core IBM POWER9 Central Processing Unit (CPU) @ 2.4GHz, and 4 16 GB NVIDIA V100 Graphical Processing units (GPUs) for a period of about 2.5 days \cite{hal_uiuc}. 

The progressive GANs (ProGANs) were trained using the original implementation provided by Karras \textit{et al.} \cite{progan_code}. The default settings for the training parameters were employed in this study. As implemented elsewhere \cite{iagan_nyu, iagan}, the default latent space dimensionality of 512 was maintained. The training was performed on a system with an Intel Xeon E5-2620v4 CPU @ 2.1 GHz and 4 NVIDIA TITAN X GPUs. The algorithms are implemented in Python 3.6/Tensorflow 1.14.
Random i.i.d. draws from the generative models are shown in \autoref{fig:samples}, along with the real samples from the training dataset. The Fr\'echet Inception Distance (FID) scores for the invertible generative model and  the state-of-the-art progressive GAN were calculated by use of the official Python implementation \cite{fid_code} and are shown in \autoref{tab:fid}. Further details regarding FID scores can be found in the literature \cite{fid, fid_desc}.

$\quad$

\begin{minipage}[!t]{0.45\linewidth}
\noindent
\resizebox{0.9\linewidth}{!}{
\begin{tabular}{@{}ccc@{}}
\toprule
Dataset      & ProGAN & INN   \\ \midrule
Knee & 22.72  & 75.06 \\
Brain  & 10.67  & 82.41 \\ \bottomrule
\end{tabular}
}
\captionsetup{justification=centering}
\captionof{table}{FID scores of the generative models. A lower FID is correlated with improved visual quality of generated images \cite{fid}.}
\label{tab:fid}
\end{minipage}
\begin{minipage}[!t]{0.5\linewidth}
\centering
\centerline{\includegraphics[width=\linewidth]{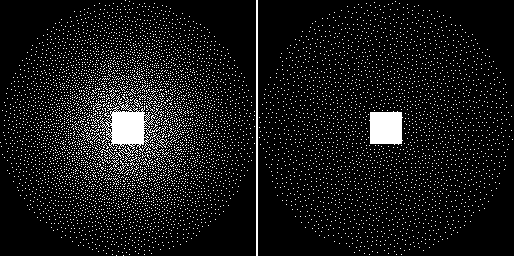}}
\captionof{figure}{8-fold (left) and 20-fold (right) undersampling masks}
\label{fig:undersampling_masks}
\end{minipage}


\subsection{Image reconstruction}
In order to assess the effect of each of the two regularization parameters associated with the INN-based reconstruction method as well as debiasing, an ablation study was performed. Simulated measurements corresponding to the R = 20 undersampling ratio were used. Also, complex i.i.d Gaussian noise was added to the measurements, such that the per-pixel SNR of the measurements (with respect to signal power) was 20 dB. The images were reconstructed with (i) no regularization, (ii) only the latent subspace projection, (iii) only the latent subspace projection followed by debiasing, (iv) only the TV penalty, (v) the latent subspace projection and the TV penalty, and (vi) the latent subspace projection with the TV penalty followed by debiasing.

Next, a coronal knee image was reconstructed from {simulated} fully sampled, noiseless $k$-space data so that the forward operator is bijective, and the loss decay was analyzed as the iterative optimization progresses. For analyzing how accurate our approach gets to actually solving the inverse problem corresponding to the measurement model in \autoref{eqn:meas_model}, a knee image $\tilde{\f} \in S_k \cap T_{\nu}$ for k = 16384 and $\norm{\tilde{\f}}_{\rm{TV}} = 922.8$, referred to as the \textit{latent-projected} image, was considered. Measurements corresponding to the $R = 8$ undersampling ratio were simulated and images were reconstructed from these noiseless measurements.

Next, the performances of the following reconstruction methods were qualitatively and quantitatively compared - (i) penalized least squares with TV regularization (PLS-TV) solved with the fast iterative shrinkage and thresholding algorithm (FISTA) \cite{fista}, (ii) the method proposed by Bora, \textit{et. al} \cite{bora}, i.e. the problem stated in \autoref{eqn:csgm}, 
with a ProGAN \cite{progan} trained as described in Section \ref{sec:num_studies}B as the generative model (henceforth referred to as CSGM-GAN), 
(iii) Image-adaptive GAN-based reconstruction with TV regularization described in equation \autoref{eqn:iagan} (IAGAN-TV) \cite{iagan, iagan_sayantan}, and (iv) INN-based reconstruction using latent space projection and TV regularization, described in equation \autoref{eqn:csgm_inn} (INN Proj. TV).
{For the simulation study,} coronal knee and axial brain images were reconstructed from {simulated} measurements corresponding to the R = 8 and R = 20 undersampling ratios for this comparison. This was done with noiseless measurements, as well as measurements with i.i.d Gaussian noise with 20 dB per-pixel SNR. 

Next, the approaches described above were employed to reconstruct anomalous pediatric brain images from 8-fold {simulated} undersampled measurements with 20 dB SNR \cite{paedomri_dataset}. The generative models used to reconstruct the pediatric brain image were trained on axial adult brain images from the previously described NYU fastMRI initiative database. 

Finally, for the emulated experimental study, image reconstruction was performed from four-fold retrospectively undersampled emulated single-coil measurements. The image estimates were compared to IFFT-based reconstructions from fully sampled measurements.

\section{Results and evaluation}\label{sec:results}
\begin{figure}[!t]
\centerline{\includegraphics[width=\linewidth]{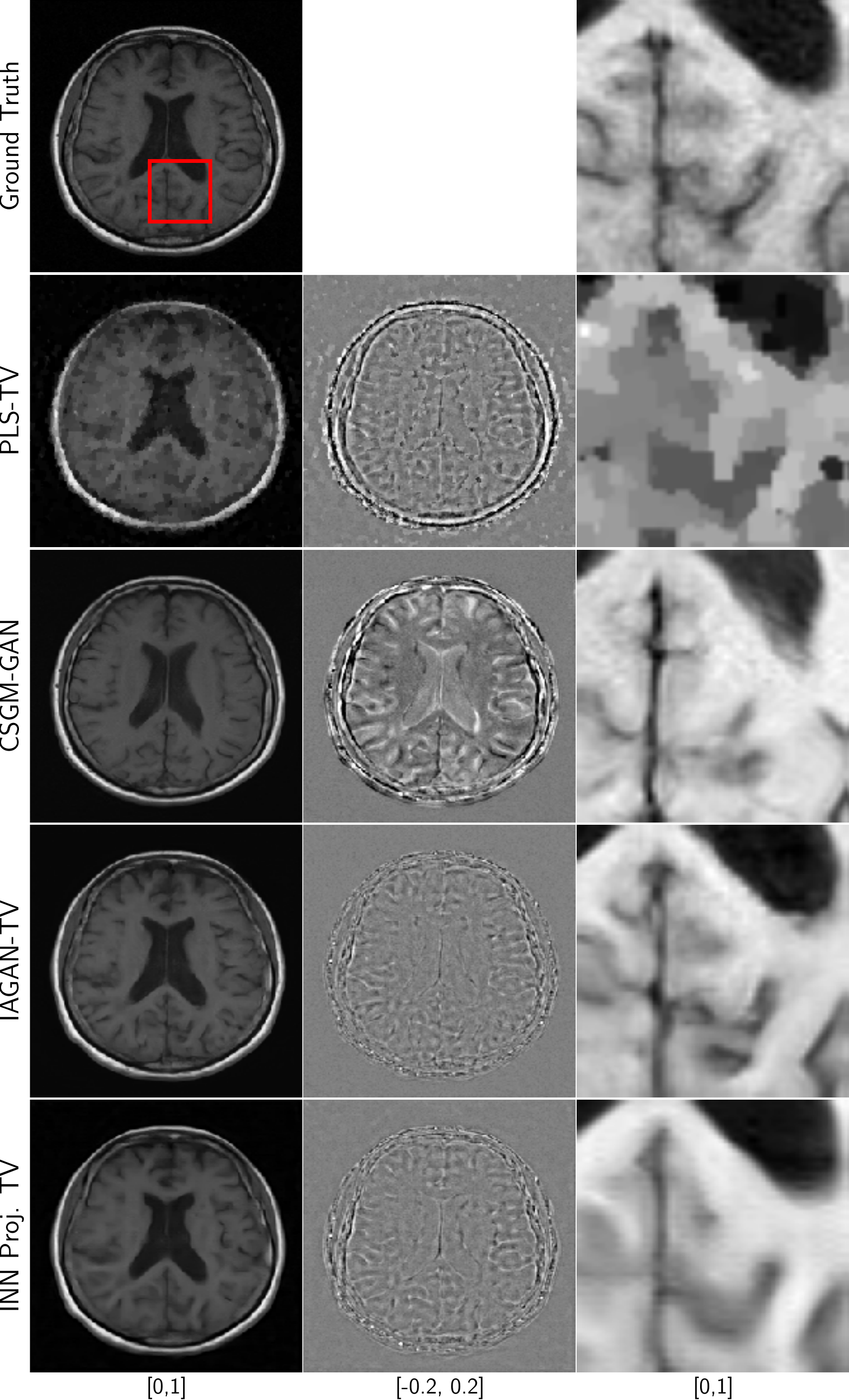}}
\caption{Ground truth, difference plots and reconstruction results for an axial T1 weighted brain image, with 20-fold undersampling and 20 dB measurement SNR. The RMSE and SSIM values are displayed in \autoref{tab:mse_ssim}.}
\label{fig:brain20x}
\end{figure}
\begin{figure}[!t]
\centerline{\includegraphics[width=\linewidth]{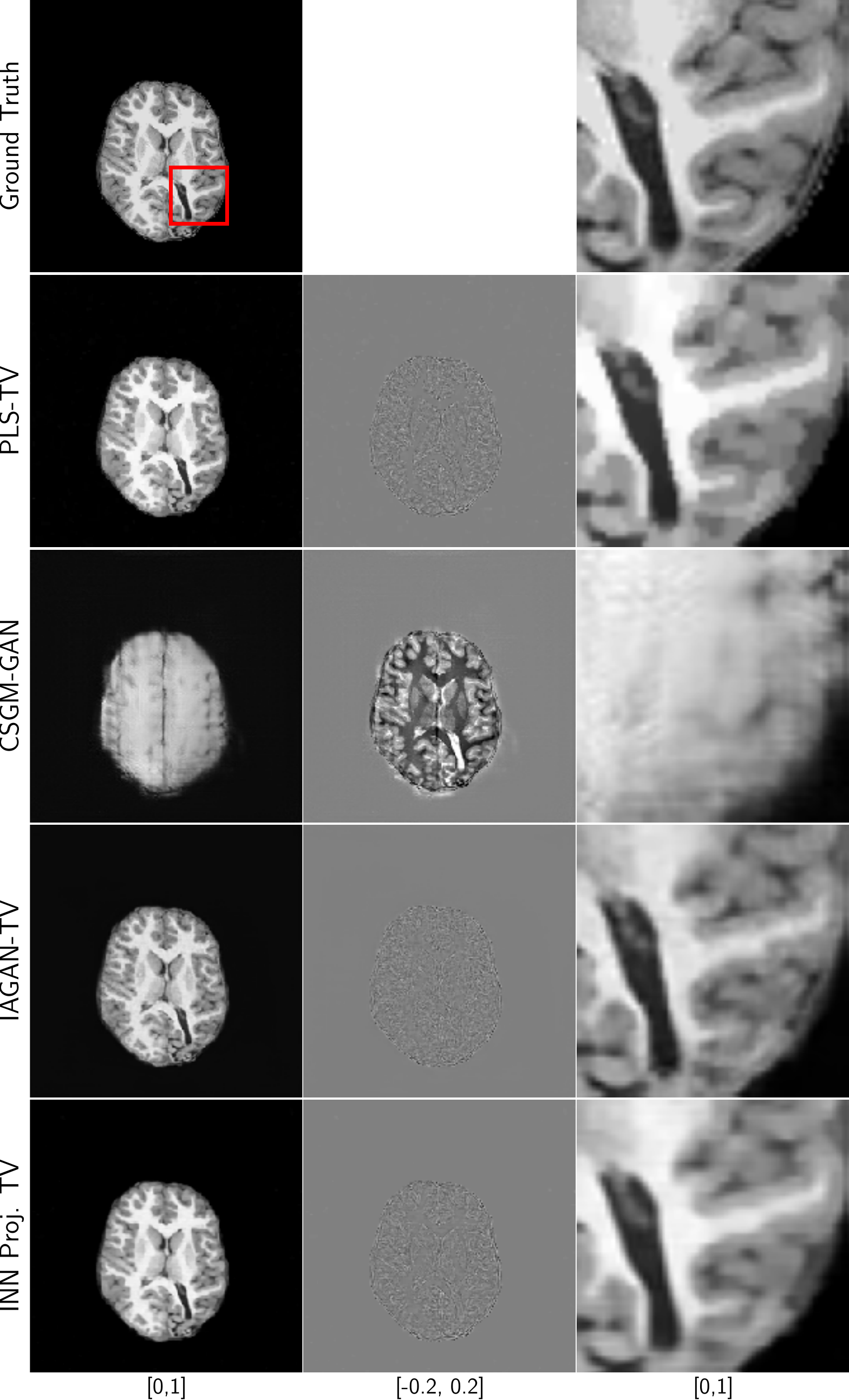}}
\caption{Ground truth, difference plots and reconstruction results for an axial T1 weighted pediatric brain image with anomaly, with 8-fold undersampling and 20 dB measurement SNR.   The RMSE and SSIM values are displayed in \autoref{tab:mse_ssim2}.}
\label{fig:brainood8x}
\end{figure}
\begin{figure}[!t]
\centerline{\includegraphics[width=\linewidth]{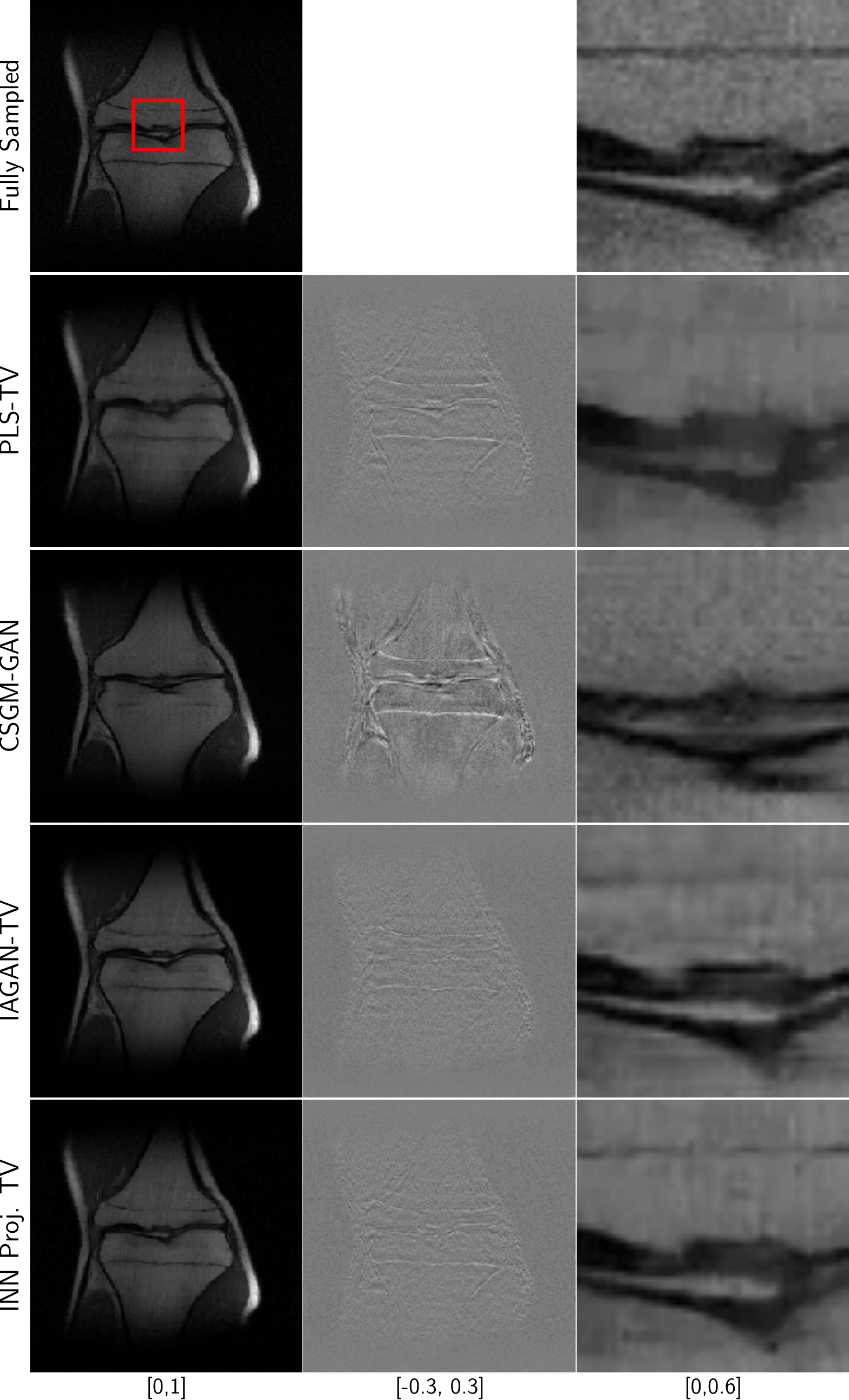}}
\caption{The absolute value of coronal PD weighted knee images reconstructed from emulated single-coil measurements with Cartesian four-fold retrospective undersampling. The RMSE and SSIM values are displayed in \autoref{tab:mse_ssim2}.}
\label{fig:experimental_4x}
\end{figure}
This section is organized as follows. First, the results of the ablation study are described. The results for the fully sampled, noiseless reconstruction and the associated RMSE and SSIM values and convergence analysis is deferred to the supplementary section. This is followed by the results of the stylized study where the object lies in $S_k \cap T_{\nu}$, and thus perfectly satisfies the measurement model. Next, the results for the test images that do not lie in $S_k \cap T_{\nu}$ are shown, after which the RMSE and SSIM comparisons, {statistical significance tests,} and the bias-variance trade-off calculations are described. 

\subsection{The ablation study}
The results of the ablation study are shown in \autoref{fig:proj_comparison}. It was observed that the best RMSE performance was achieved by a combination of latent subspace projection and TV penalty, without debiasing. Hence, in the rest of the manuscript, results using this particular combination of regularization parameters will be described. It should be noted, however, that a combination of latent subspace projection and TV penalty followed by debiasing was able to improve upon the performance of the chosen method if early stopping was performed during debiasing. However, this adds an additional tunable parameter. In the interest of simplicity, the use of debiasing was avoided. 


\subsection{Reconstruction of latent-projected images}
Estimates of 20 latent-projected images were obtained from noiseless, 8-fold undersampled measurements. One of the reconstructed images is shown along with the ground truth in the supplementary section. The mean and standard deviation of RMSE and SSIM values obtained over the ensemble were $0.0046 \pm 0.0007$ and $0.9956 \pm 0.0012$. 
This indicates that the measurement model was not exactly inverted. Here, although the proposed method performs worse than the noiseless, fully sampled case, it performs significantly better as compared to the noiseless undersampled case with the test data. This suggests that the inverse problem when restricted to $S_k \cap T_\nu$ is less ill-conditioned than the case where $\tilde{\f} \not\in S_k \cap T_\nu$.

\subsection{Reconstruction of knee and brain test images}
\subsubsection{{Simulation study: Reconstruction from noiseless undersampled measurements}}
The RMSE and SSIM evaluation metrics for the test knee and brain images reconstructed from noiseless undersampled data, and the corresponding images are displayed in the supplementary section. 

\subsubsection{{Simulation study: Reconstruction from undersampled measurements with 20 dB measurement SNR}}
Figures \ref{fig:knee8x} and \ref{fig:knee20x} display reconstructed images of a coronal knee {test} image from 8-fold and 20-fold {noisy} undersampled measurements respectively. One key observation is that for 8-fold subsampling, all algorithms except for CSGM-GAN performed well, in terms of RMSE and SSIM. This was because the 8-fold variable density Poisson disc undersampling mask is designed in order to keep the low frequency information intact, and randomly sample only the high frequency information with a variable density. It should be noted that due to the representation error, the CSGM-GAN reconstruction retained highly realistic features, some of which, were false. 
Further, it should be noted that the IAGAN-TV and the INN-based method seem to have performed the best in terms of recovering the finer features of the image. As shown in \autoref{fig:knee20x}, for 20-fold undersampling, it was seen that the PLS-TV reconstruction has characteristic smoothing artifacts due to the TV regularization. Choosing lower regularization values led to noisier images, as shown in the supplementary section.

Similar observations can be made for the results of reconstruction of an axial brain image from 8-fold and 20-fold undersampled measurements, as shown in \autoref{fig:brain8x} and \autoref{fig:brain20x}, respectively. In addition, it should be noted that for the 8-fold undersampling case, some of the finer features, such as the folds in the brain, are difficult to recover using PLS-TV, but were successfully recovered with both IAGAN-TV and the INN-based reconstruction. For the 20-fold undersampling case, all the methods face challenges in recovering finer features such as the folds of the brain, and may produce oversmoothened features or even realistic hallucinations. Finally, the results for the reconstruction of the pediatric brain image are shown in \autoref{fig:brainood8x}. Here, the out-of-distribution image was accurately recovered by the IAGAN-TV and the proposed method, but not in the case of CSGM-GAN. {Here, the poor performance of CSGM-GAN could also be due to domain shifts unrelated to anatomical features.}

\subsubsection{Emulated experimental study}
The absolute value of the image reconstructed by use of the proposed INN-based method from four-fold retrospectively undersampled emulated single-coil measurements is shown in \autoref{fig:experimental_4x}, along with the IFFT-based reconstruction from fully sampled data and difference plots. Here, we see that the proposed method demonstrates superior performance among all the examined methods.


\subsubsection{Root Mean square error and structural similarity}
Root mean-squared error (RMSE) and structural similarity (SSIM) index values over an ensemble of 50 test images of each category described above were calculated. Ensemble mean and standard deviation of these values are displayed in Tables \ref{tab:mse_ssim} and \ref{tab:mse_ssim2}. It can be noted that, across several image categories, the performance of the IAGAN-TV and the proposed method was comparable and the best among all the methods compared, although IAGAN-TV outperformed the proposed method for some of the image categories by a small margin. For the emulated experimental study, RMSE and SSIM values were computed with respect to the fully-sampled IFFT-based reconstruction. The performance of the proposed method outperformed all other examined methods for this study.

The statistical significance of the differences between the reconstruction methods was tested using the one way repeated measures ANOVA test, followed by post-hoc paired samples $t$-tests between pairs of algorithms, with the Bonferroni correction. Since the metrics obtained from CSGM-GAN violated some of the assumptions of the ANOVA test, it was left out of the statistical significance study. It was observed that IAGAN-TV and the proposed approach are both statistically significantly better than PLS-TV (with $p\text{-value} < 10^{-17}$ for the in-distribution images, $p\text{-value} < 10^{-12}$ for the out-of-distribution images, and $p\text{-value} < 10^{-6}$ for the emulated experimental study). For the in-distribution images in the simulation study, there is a small but statistically significant difference between the performance of IAGAN-TV and the proposed method, with IAGAN performing better (with $p\text{-value} < 10^{-5}$). However, for the out-of-distribution image, no statistically significant difference was observed between the two approaches. For the emulated experimental study, the proposed method significantly outperformed IAGAN-TV (with $p\text{-value} < 10^{-8}$).


\begin{table*}[!h]
\centering
\caption{Comparison of RMSE and SSIM for different algorithms for undersampled data with 20 dB measurement SNR, for the simulation study, computed on an ensemble of 50 images. The values outside the parentheses denote the ensemble mean values of the metric, where as the values inside the parentheses denote the standard deviation (SD) of the metric.}
\label{tab:mse_ssim}
\resizebox{0.95\linewidth}{!}{%
\begin{tabular}{cccccccccc}
\toprule
Algorithm    & {} & \multicolumn{2}{c}{Knee (in dist.) 8x} & \multicolumn{2}{c}{Knee (in dist.) 20x} & \multicolumn{2}{c}{Brain (in dist.) 8x} & \multicolumn{2}{c}{Brain (in dist.) 20x} \\
             & {} &    RMSE mean  & SSIM mean  &     RMSE mean  & SSIM mean  &     RMSE mean  & SSIM mean  &      RMSE mean  & SSIM mean  \\
             & {} &    (RMSE SD) & (SSIM SD) &      (RMSE SD) &  (SSIM SD) &   (RMSE SD) & (SSIM SD) &     (RMSE SD) &  (SSIM SD) \\
\midrule
\multirow{2}{*}{PLS-TV}         & \multirow{8}{*}{{}} &             0.0122 &         0.9736 &              0.0178 &         0.9556 &              0.0228 &        0.9609 &               0.0473 &         0.8798 \\
                                & \multirow{7}{*}{{}} &           (0.0033) &       (0.0108) &            (0.0050) &       (0.0172) &            (0.0033) &      (0.0093) &             (0.0098) &       (0.0337) \\
\cline{1-10}                                                                                                                                                                                                        
\multirow{2}{*}{CSGM-GAN}       & \multirow{6}{*}{{}} &             0.0381 &         0.8808 &              0.0389 &         0.8753 &              0.0721 &        0.8174 &               0.0725 &         0.8153 \\
                                & \multirow{5}{*}{{}} &           (0.0157) &       (0.0485) &            (0.0154) &       (0.0470) &            (0.0318) &      (0.0633) &             (0.0249) &       (0.0598) \\
\cline{1-10}                                                                                                                                                                                                       
    \multirow{2}{*}{IAGAN-TV}       & \multirow{4}{*}{{}} &        \bf{0.0099} &    \bf{0.9844} &         \bf{0.0140} &    \bf{0.9705} &         \bf{0.0148} &   \bf{0.9794} &          \bf{0.0246} &    \bf{0.9483} \\
                                    & \multirow{3}{*}{{}} &           (0.0026) &       (0.0064) &            (0.0041) &       (0.0135) &            (0.0024) &      (0.0061) &             (0.0043) &       (0.0146) \\
    \cline{1-10}                                                                                                                                                                                                       
    \multirow{2}{*}{INN Proj. TV}   & \multirow{2}{*}{{}} &             0.0102 &         0.9829 &              0.0147 &         0.9678 &              0.0163 &        0.9723 &               0.0262 &         0.9414 \\
                                                     & {} &           (0.0027) &       (0.0070) &            (0.0042) &       (0.0137) &            (0.0028) &      (0.0086) &             (0.0049) &       (0.0177) \\
    \bottomrule
    \end{tabular}}
    
    \end{table*}
    
\begin{table}[!t]
\centering
\caption{Comparison of RMSE and SSIM for different algorithms over an ensemble of 50 out-of-distribution brain images for the simulation study and 40 knee images for the emulated experimental study. The values inside the parentheses denote the standard deviation (SD) of the metric.}
\label{tab:mse_ssim2}
\resizebox{\linewidth}{!}{%
\begin{tabular}{ccccc}
\toprule
Algorithm    & \multicolumn{2}{c}{Brain (out of dist.)} & \multicolumn{2}{c}{Emulated experimental} \\
             &    RMSE mean  & SSIM mean  &     RMSE mean  & SSIM mean  \\
             &    (RMSE SD) & (SSIM SD) &      (RMSE SD) &  (SSIM SD) \\
\midrule
\multirow{2}{*}{PLS-TV}         & \multirow{8}{*}{{}}        0.0124 &         0.9813 &      0.0148 &         0.9151 \\
                                & \multirow{7}{*}{{}}      (0.0012) &       (0.0046) &    (0.0046) &       (0.0459) \\
\cline{1-5}                                                                                                                       
\multirow{2}{*}{CSGM-GAN}       & \multirow{6}{*}{{}}        0.0516 &         0.7932 &      0.0283 &         0.8815 \\
                                & \multirow{5}{*}{{}}      (0.0088) &        (0.022) &    (0.0072) &        (0.0455) \\
\cline{1-5}                                                                                                                      
\multirow{2}{*}{IAGAN-TV}       & \multirow{4}{*}{{}}        0.0119 &    \bf{0.9847} &      0.0142 &         0.9098 \\
                                & \multirow{3}{*}{{}}      (0.0013) &       (0.0041) &    (0.0049) &       (0.0590) \\
\cline{1-5}                                                                                                                       
\multirow{2}{*}{INN Proj. TV}   & \multirow{2}{*}{{}}   \bf{0.0118} &         0.9846 & \bf{0.0135} &     \bf{0.9320} \\
                                                 & {}      (0.0011) &       (0.0038) &    (0.0044) &       (0.0360) \\
\bottomrule
\end{tabular}}

\end{table}

\subsubsection{Bias-Variance tradeoff}
\begin{figure}[!t]
\centerline{\includegraphics[width=\linewidth]{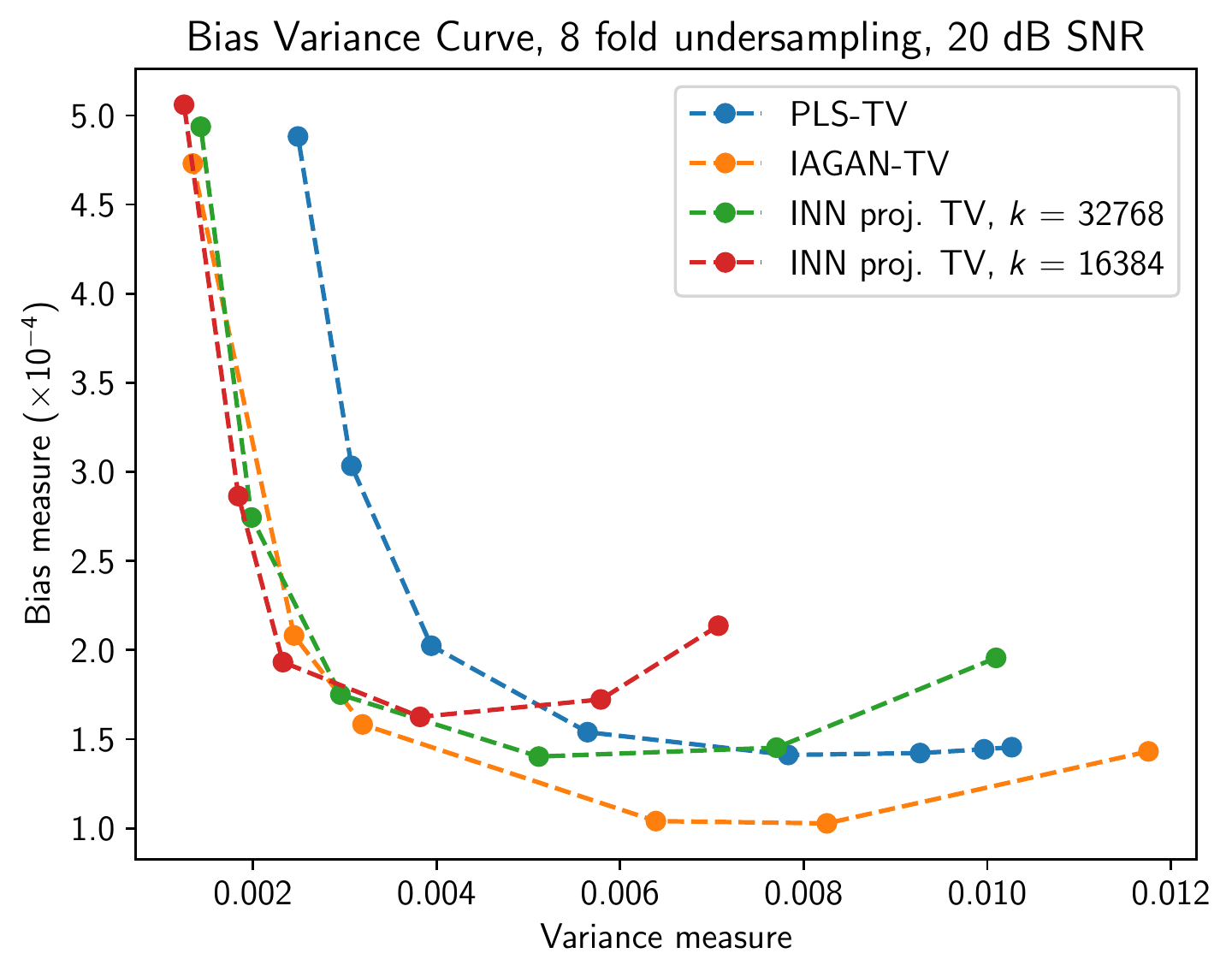}}
\caption{Bias variance tradeoff analysis for 8-fold undersampling comparing PLS-TV, IAGAN-TV and the proposed method, while sweeping $\mu$.}
\label{fig:bv_8x}
\end{figure}

Although the evaluation of perceptual quality and quantitative evaluation in terms of RMSE and SSIM indicate the superiority of the INN-based reconstruction as compared to more traditional approaches, a task-based assessment of reconstruction algorithms is necessary to determine the superiority of one reconstruction algorithm to the other \cite{barrett}. However, such a detailed task-based assessment of generative models based reconstruction algorithms is a substantial task in itself, and remains a topic for future study. Here, an analysis of the bias-variance trade-off is provided.



As described in equation \autoref{eqn:csgm_inn}, the INN-based reconstruction method involves the use of two explicit regularization parameters - (i) $k$, the dimensionality of the latent subspace containing the most important $\z$ components, and (ii) $\mu$, the weight of the TV regularization. In the presented analysis, for a fixed value of $k$, $\mu$ was swept to obtain different values of bias. This entire procedure was repeated for another value of $k$. The results of both parts were compared with PLS-TV and IAGAN-TV, where the TV regularization weight was swept.

\begin{figure}[!t]
\centerline{\includegraphics[width=\linewidth]{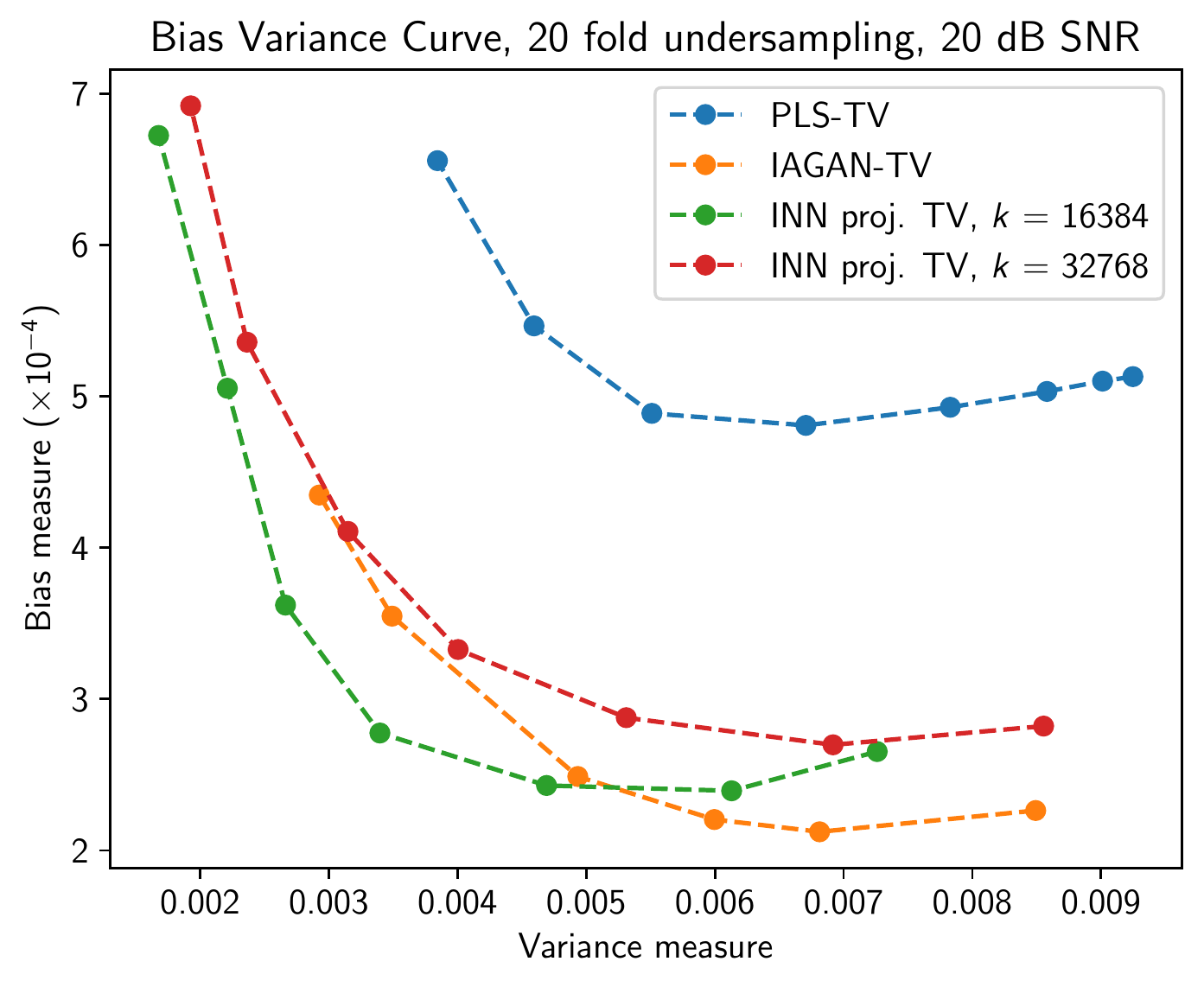}}
\caption{Bias variance tradeoff analysis for 20-fold undersampling comparing PLS-TV, IAGAN-TV and the proposed method, while sweeping $\mu$. \vspace{-0.2mm}}
\label{fig:bv_20x}
\end{figure}

Bias-variance analysis was performed on images reconstructed from {simulated} measurements corresponding to both 8-fold and 20-fold undersampling patterns, with 20 dB measurement SNR.
The ground truth used for this study was an image from the fastMRI knee dataset. Stylized, simulated undersampled single-coil MRI measurements were used.
A dataset of reconstructed images $\{\hat{f}^{(i)}\}_{i=1}^d$ from measurements with $d = 100$ independent noise realizations was considered for every regularization setting. The bias $\vec{b}$ and the variance $\sigma_i$ of a pixel $i$ were calculated as:
\begin{align}
    \bf{b} =&~ \frac{1}{d}\sum_{i=1}^d \f^{(i)} - \tilde{\f}\\
    \sigma_j^2 =&~ \frac{1}{d-1} \left(\f^{(i)}_j - \frac{1}{d}\sum_{i=1}^d \f^{(i)}_j \right)^2,
\end{align}
where $\tilde{\f}$ is the ground truth image. As a summary measure, {the average squared bias $\frac{1}{n}\norm{\bf{b}}^2_2$ versus the average variance $\frac{1}{n}\sum_{j=1}^n \sigma_j^2$ was plotted.} Figures \ref{fig:bv_8x} and \ref{fig:bv_20x} show the bias-variance curves for 8 and 20-fold undersampling, respectively.

As can be seen, the bias and variance curves for the INN-based method lie below the curves for PLS-TV, which is indicative of superior performance over a range of regularization values. This also indicates that, while the transition from an over-smoothed image to a noisy image is such that intermediate images could be both noisy and oversmoothed, this trade-off is better for the proposed reconstruction approach. 

\section{Discussion and conclusion}\label{sec:discussion}
Due to the design of the measurement operator in the case of variable density Poisson disc undersampling, PLS-TV outperformed CSGM-GAN in most cases in terms of MSE and SSIM. The discrepancies in the image estimate from CSGM-GAN are also visually evident from the difference plot. This is consistent with the observation made by Bora \textit{et. al} \cite{bora}, where they report that CSGM-GAN outperformed sparsity-based methods (with respect to MSE) only in the cases of severe undersampling. Here, it can be seen that if the measurement operator is well designed, it is possible to have formally severe undersampling scenarios where the PLS-TV outperforms CSGM-GAN. Moreover, it was observed that that the INN-based methods, as well as IAGAN-TV were successful in removing the plausible but false features that could be present in images reconstructed by CSGM-GAN. 

Better trained networks give better performance, since they impose a better generative prior - a fact that was also observed when testing reconstruction using INNs trained with poorer hyperparameter settings. The current state-of-the-art GANs possess superior generative performance compared to state-of-the-art for invertible generative models, as evidenced by literature \cite{glow, flowpp, progan} as well as the FID scores calculated in \autoref{tab:fid}. This explains why in several cases, IAGAN-TV outperformed the proposed method. However, one key thing to note here is that for the INN-based reconstruction, the parameters of the network were \textit{not} adapted, and still a performance very close to IAGAN-TV was achieved. Also, the IAGAN-TV performance was achieved using early stopping.

Moreover, the number of parameters that need to be optimized for the IAGAN-TV would increase as the complexity of GANs increases, where for the INN, optimization over only the latent-space vector was needed in order to achieve comparable performance.
For instance, in this study, IAGAN-TV requires optimization over more than 23 million parameters whereas the proposed approach require optimization over less than 33000 parameters for image reconstruction from simulated measurements, and around 65000 parameters for the experimental measurements. Finally, the IAGAN-TV optimization was carried out by initializing with the CSGM-GAN solution, which itself required about 10 independent random restarts to achieve a reasonable reconstruction. Including good initialization via CSGM-GAN and the several random initializations needed thereof, the net runtime for IAGAN-TV is around 120 minutes on a single NVIDIA 1080 Ti GPU, whereas the proposed method requires around 20 minutes.

It is important to note how the studies conducted here relate to a real experimental MRI scenario. The forward operator used in the iterative reconstruction in all our studies was the discrete Fourier transform followed by undersampling in the Fourier domain based on a binary mask. This is a stylized simulation and does not reflect all the complexities of the true MRI forward model, such as the single-coil sensitivity. For the in-distribution simulation studies, the ground truth images were real-valued floating point numbers. For the out-of-distribution simulation studies, however, the ground truth images were originally real-valued, skull-stripped and compressed to 8 bit integers. The noise model for the simulated study was iid Gaussian. All these factors may simplify the image reconstruction task substantially. For the emulated experimental study, the IFFT of the fully sampled $k$-space corresponds to complex-valued floating point numbers and has non-trivial phase variations. Since image reconstruction was performed directly from the emulated undersampled single-coil measurements, the implicit noise model is expected to be more realistic than the simulation studies. The undersampling, however, was retrospective and the measurement data were generated as a linear combination of responses from raw multi-coil data.



In conclusion, a new method of image reconstruction from incomplete measurements using invertible generative priors was proposed, based on a novel regularization strategy for INNs with a multiscale architecture. This method was evaluated and compared with other competing methods on the problem of estimating images from simulated {and emulated experimental} undersampled MRI measurements. Some important extensions of this work include {comparisons with regularizaton methods adapted to continuous signal variation such as total generalized variation (TGV) \cite{tgv},}  developing strategies for {multi-coil MRI} and 3D image reconstruction, as well as a task-based evaluation of generative model-based image reconstruction methods. Another interesting avenue for future research is examining different strategies for penalizing the latent vector, such as constraining the latent vector to lie on a surface on which a random latent vector concentrates \cite{sr_stylegan}.
{
\bibliographystyle{IEEEtran}
\bibliography{csmri}
}


%




\ifCLASSOPTIONcaptionsoff
  \newpage
\fi

\begin{IEEEbiography}[{\includegraphics[width=1in,height=1.25in,clip,keepaspectratio]{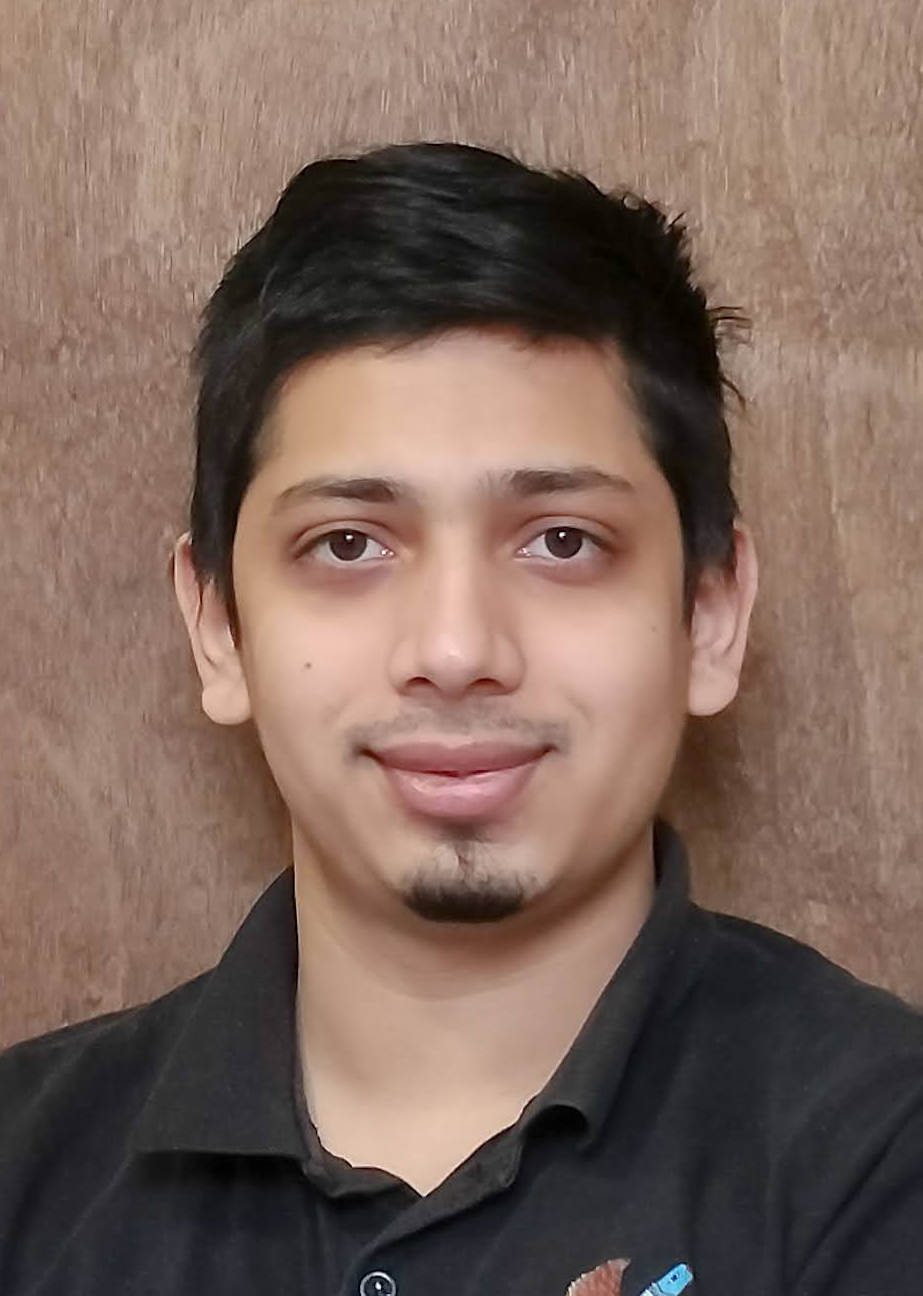}}]{Varun A. Kelkar}
is a Ph.D. candidate in the Department of Electrical and Computer Engineering, University of Illinois at Urbana-Champaign, IL, USA. He received the M.S. degree in Electrical and Computer Engineering from UIUC in 2019, and the B.Tech. degree in Engineering Physics from the Indian Institute of Technology Madras, TN, India in 2017. His research interests include computational imaging, inverse problems, signal processing, optics and machine learning. He is a member of the SPIE, and was a recipient of the SPIE Optics and Photonics Education Scholarship in 2019.
    
\end{IEEEbiography}
    
\begin{IEEEbiography}[{\includegraphics[width=1in,height=1.25in,clip,keepaspectratio]{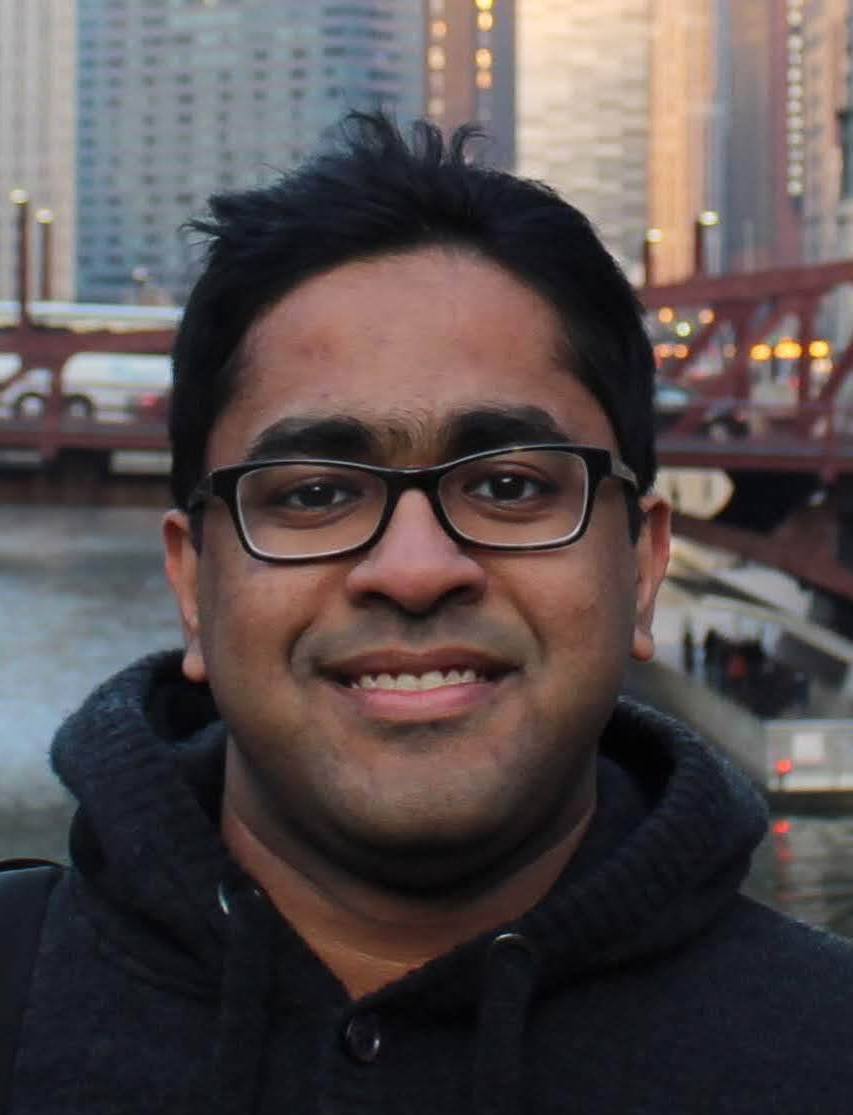}}]{Sayantan Bhadra}
received the B.E. degree in Electronics and Telecommunication Engineering from Jadavpur University, West Bengal, India, in 2016. He is currently working toward the Ph.D degree in Computer Science and Engineering at the Washington University in St. Louis, MO, USA. He is also a visiting research scholar in the Computational Imaging Science Laboratory, Department of Bioengineering, University of Illinois at Urbana-Champaign, IL, USA. His research interests include image reconstruction and machine learning for medical imaging applications.
\end{IEEEbiography}
    
\begin{IEEEbiography}[{\includegraphics[width=1in,height=1.25in,clip,keepaspectratio]{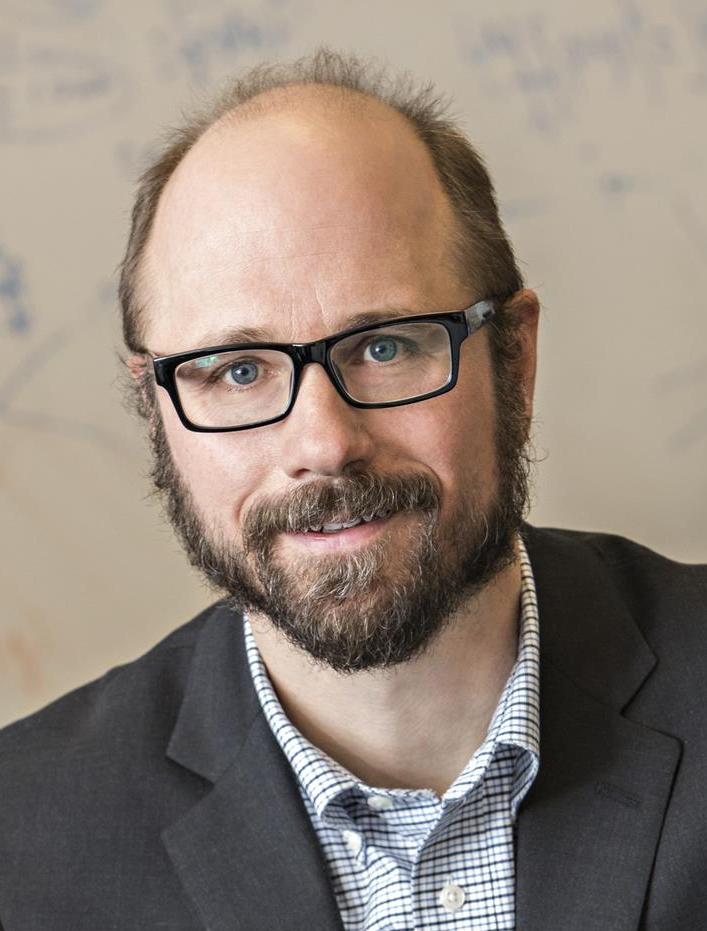}}]{Mark A. Anastasio}
is the Donald Biggar Willett Professor in Engineering and the Head of the Department of Bioengineering UIUC. Before joining UIUC in 2019, he was a professor of biomedical engineering and the founding director of one of the nation’s first stand-alone PhD programs in imaging science at Washington University in St. Louis. Dr. Anastasio’s research addresses the computational aspects of image formation, modern imaging science, and machine learning. He has conducted research in the fields of diffraction tomography, X-ray phase-contrast imaging, and ultrasound tomography. He is also a leading authority on photoacoustic computed tomography.  His research has been continuously funded by the NIH and NSF and he was the recipient of an NSF CAREER Award. He is a Fellow of the American Institute for Medical and Biological Engineering and the SPIE and served as the Chair of the NIH BMIT-B and EITA Study Sections.
\end{IEEEbiography}







\end{document}


%
\title{Compressible Latent-Space Invertible Networks for Generative Model-Constrained Image Reconstruction -- Supplementary Information}
%
%
%

\author{Varun A. Kelkar, Sayantan Bhadra, and Mark A. Anastasio, \IEEEmembership{Senior Member, IEEE}
\thanks{This work was supported in part by NIH Awards EB020604, EB023045, NS102213, EB028652, and NSF Award DMS1614305.}
\thanks{Varun A. Kelkar is with the Department of Electrical and Computer Engineering, University of Illinois Urbana-Champaign, Urbana, IL 61801 USA (e-mail: vak2@illinois.edu). }
\thanks{Sayantan Bhadra is with the Department of Computer Science and Engineering, Washington University in Saint Louis, Saint Louis, MO USA (e-mail: sayantanbhadra@wustl.edu).}
\thanks{Mark A. Anastasio is with the Department of Bioengineering, University of Illinois Urbana-Champaign, Urbana, IL 61801 USA (e-mail: maa@illinois.edu).}}

\markboth{IEEE Transactions on Computational Imaging,~Vol.~, No.~, October~2020}%
{Shell \MakeLowercase{\textit{et al.}}: Bare Demo of IEEEtran.cls for IEEE Journals}

\maketitle

\begin{IEEEkeywords}
Image reconstruction, compressive sensing, generative neural networks,
invertible neural networks
\end{IEEEkeywords}

%
\IEEEpeerreviewmaketitle

\section{Experiments for testing the compressibility of latent space of the INN}
As described in Section III of the manuscript, the effect of the latent-space compressibility for the INN was examined on an ensemble of 500 images from a test dataset of coronal knee images that was kept out of the INN training dataset. This was done as follows. First, for an image $\f_{\rm{orig}}$ in the ensemble, the exact latent representation $\z_{\rm{orig}}$ was computed. This can be divided into multiple sections $\z^{(1)}, \z^{(2)}, \dots, \z^{(L)}$ based on the multilevel architecture of the INN. Next, all sections from $\z^{(1)} \dots \z^{(i)}$ were progressively set to zero such that only 50\%, 25\%, 12.5\%, 6.25\% and 3.125\% of the components of $\z$ remain non-zero. For these modified latent space vectors $\z_{50\%}$, $\z_{25\%}$, $\z_{12.5\%}$, $\z_{6.25\%}$ and $\z_{3.125\%}$, the corresponding images $\f_{P} = G_{\rm{inn}}(\z_{P}),~ P = 50\%,~ 25\%,~ 12.5\%,~ 6.25\% \text{ and } 3.125\%$ were computed, and the root mean square errors (RMSEs) $\norm{\f_P - \f_{\rm{orig}}}/\sqrt{n}$ with respect to the original image $\f_{\rm{orig}}$ were calculated.

\noindent\begin{figure}
\begin{subfigure}[!t]{0.5\linewidth}
\includegraphics[width=\linewidth]{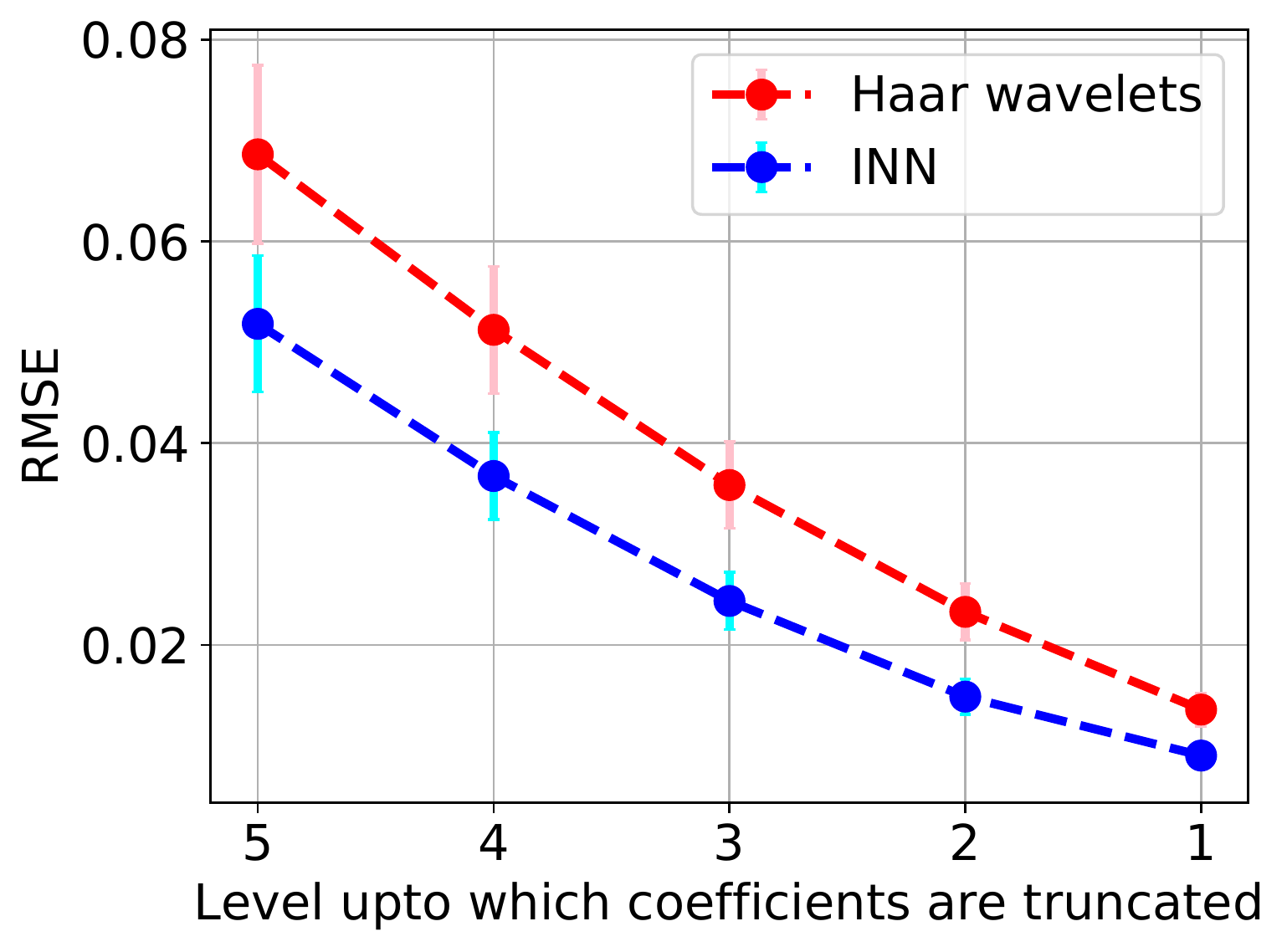}
\captionof{figure}{Truncation RMSE values}
\label{fig:rmse_truncation_haar_inn}
\end{subfigure}
\begin{subfigure}[!t]{0.5\linewidth}
\includegraphics[width=\linewidth]{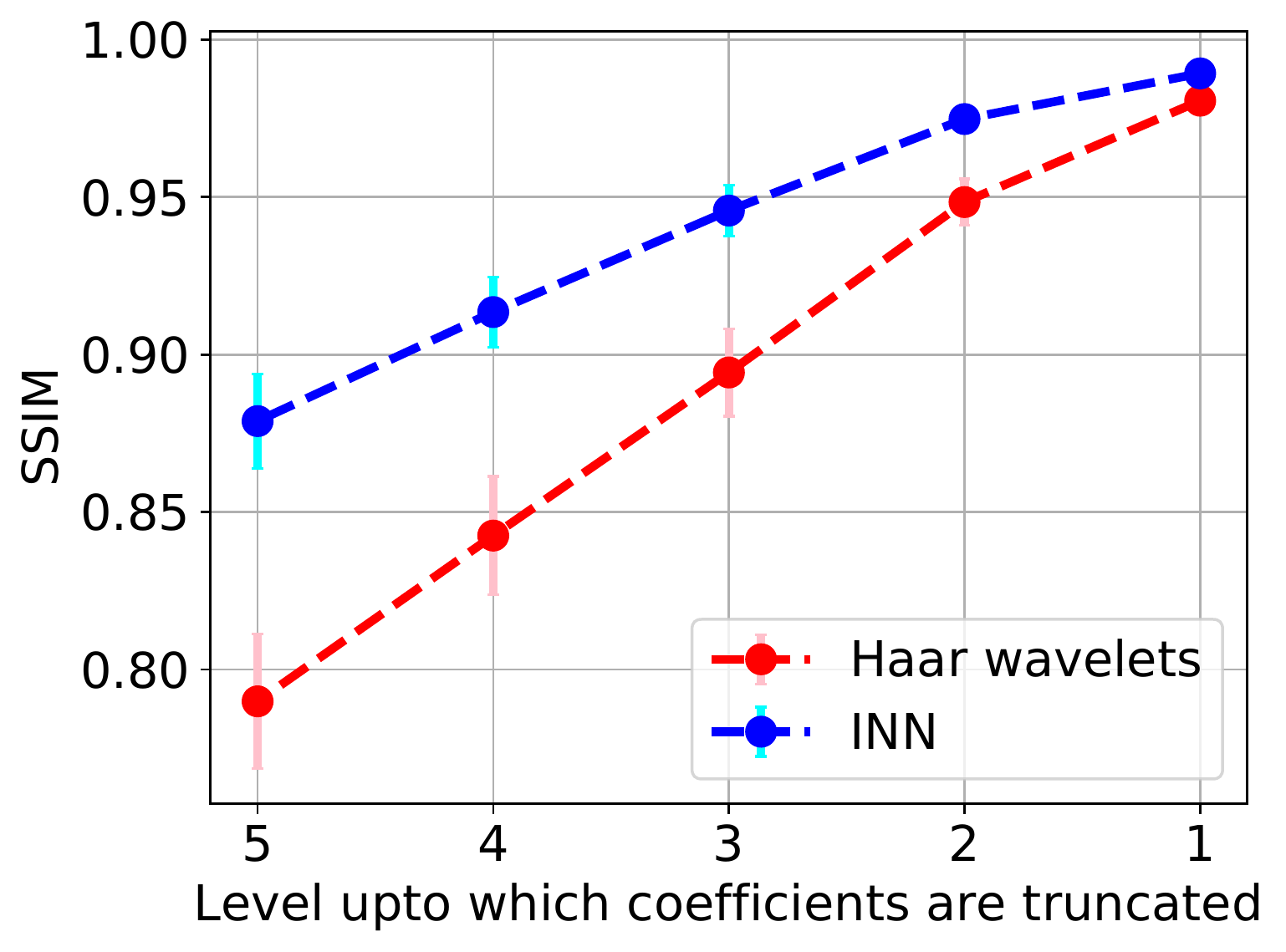}
\captionof{figure}{Truncation SSIM values}
\label{fig:ssim_truncation_haar_inn}
\end{subfigure}
\caption{Truncation RMSE and SSIM values for INN and Haar wavelet transform.}
\end{figure}

The error versus the percentage of $\z$ coefficients kept, averaged over the entire ensemble, was computed. This was then compared with the Haar wavelet transform as follows. The Haar wavelet transform with $L$ levels (same as that of the INN) was computed for each image in the ensemble. If the multilevel Haar wavelet coefficients are divided into $(\mathbf{c}^{(1)}, \mathbf{c}^{(2)}, \dots, \mathbf{c}^{(L)})$, then the $i$-th truncation level was computed by setting $\vec{c}^{(1)} \dots \vec{c}^{(i)}$ to zero. This was then compared with the $i$-th truncation level of the INN. The RMSE and SSIM truncation errors for both the INN and the Haar wavelet transform are shown in Figures \ref{fig:rmse_truncation_haar_inn} and \ref{fig:ssim_truncation_haar_inn} respectively.

\begin{figure}[!t]
\centerline{\includegraphics[width=0.6\linewidth]{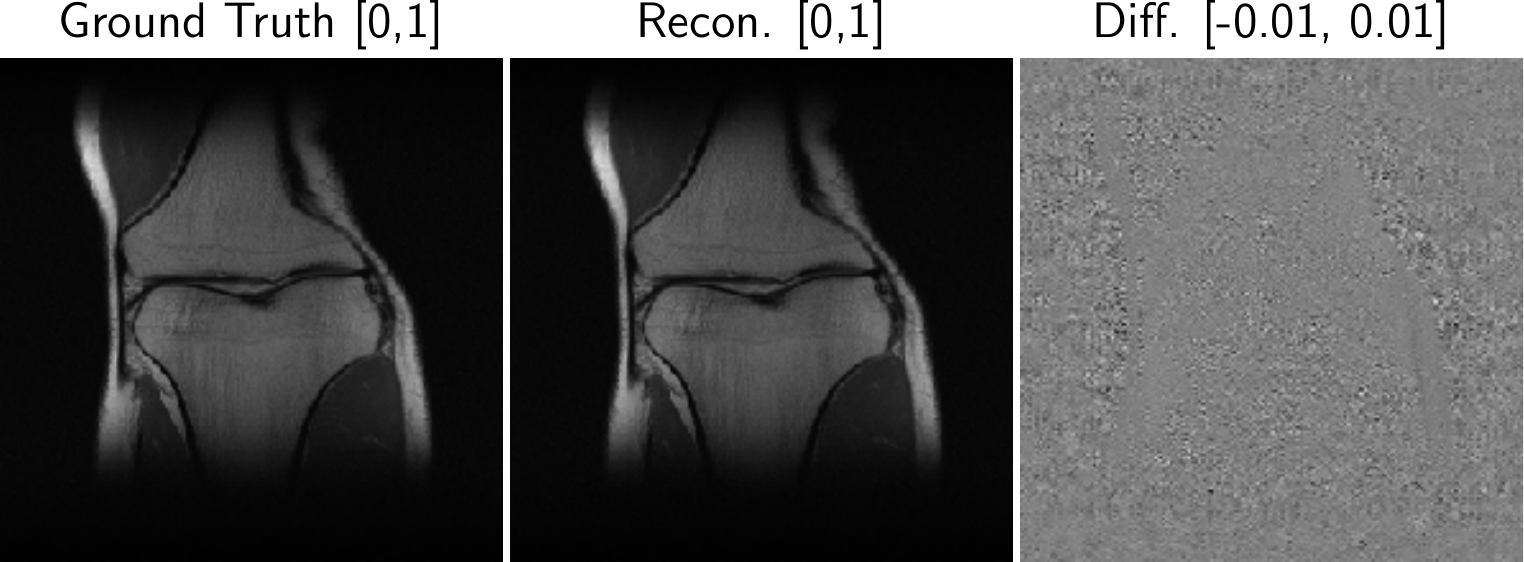}}
\caption{Ground truth, reconstruction and difference plot for reconstruction of a coronal knee image from fully sampled, noiseless measurement data}
\label{fig:fullsamp_recon}
\end{figure}
\begin{figure}
\begin{subfigure}[!t]{0.5\linewidth}
\centering
\includegraphics[width=\linewidth]{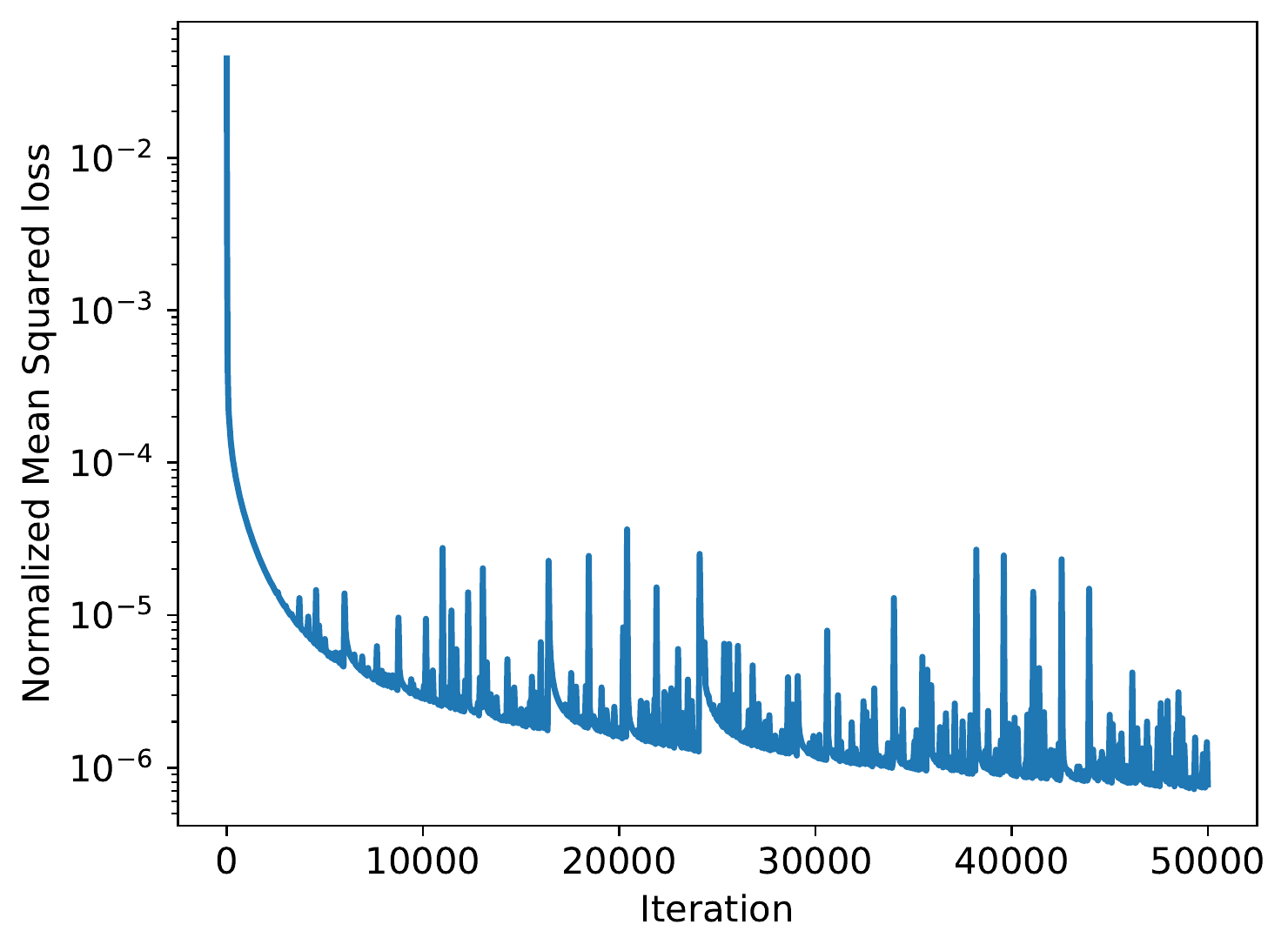}
\captionof{figure}{Normalized mean square loss plotted against the iteration for the inverse crime study.}
\label{fig:inverse_crime}
\end{subfigure}
\begin{subfigure}[!t]{0.5\linewidth}
\includegraphics[width=\linewidth]{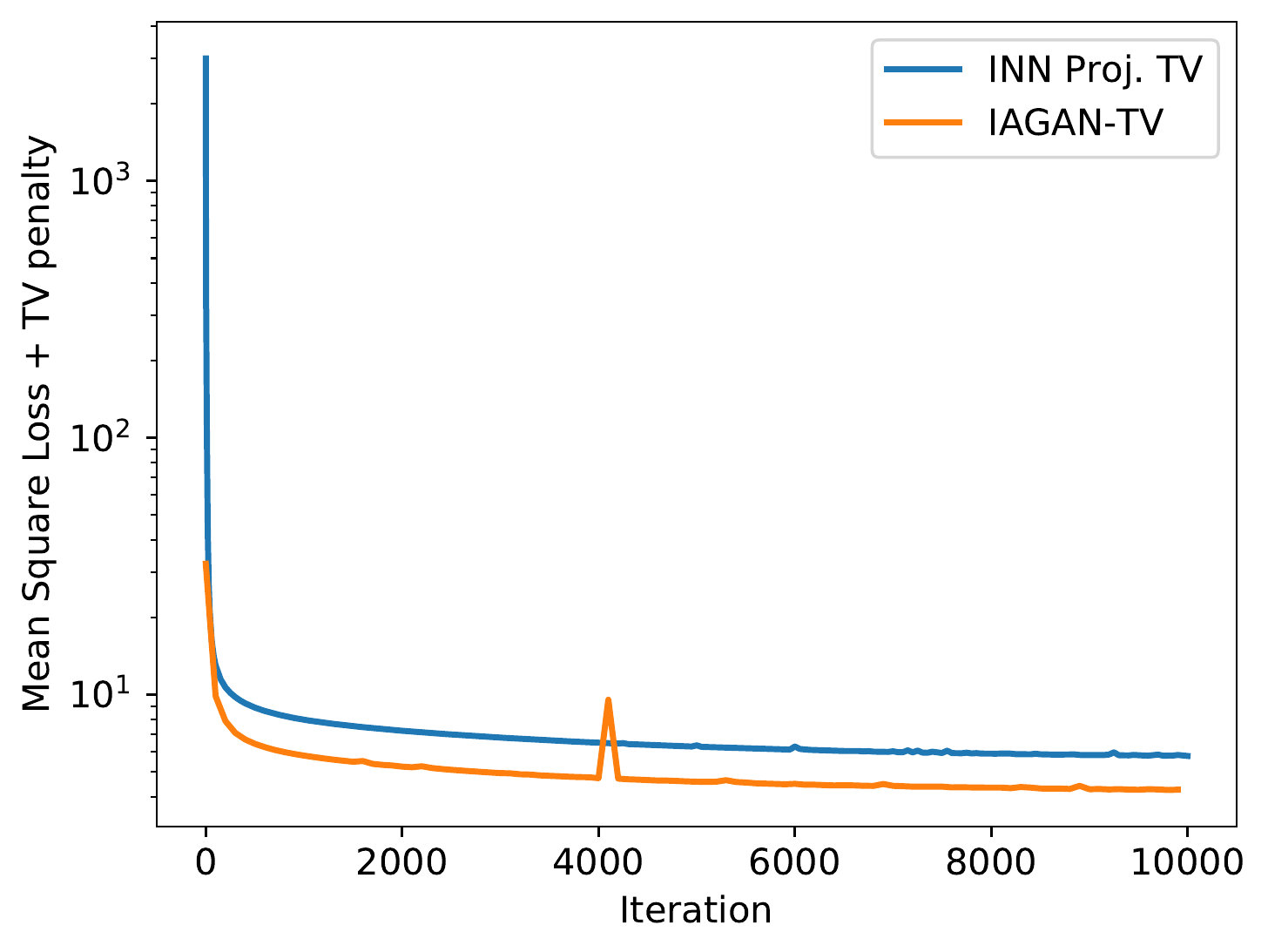}
\captionof{figure}{Mean squared loss + TV penalty versus iteration, for 8 fold noisy undersampled measurements.}
\label{fig:loss_vs_iter}
\end{subfigure}
\caption{Loss function profiles for the inverse crime study and for reconstruction from noisy undersampled simulated measurements.}
\end{figure}
\begin{figure}[!t]
\centerline{\includegraphics[width=0.6\linewidth]{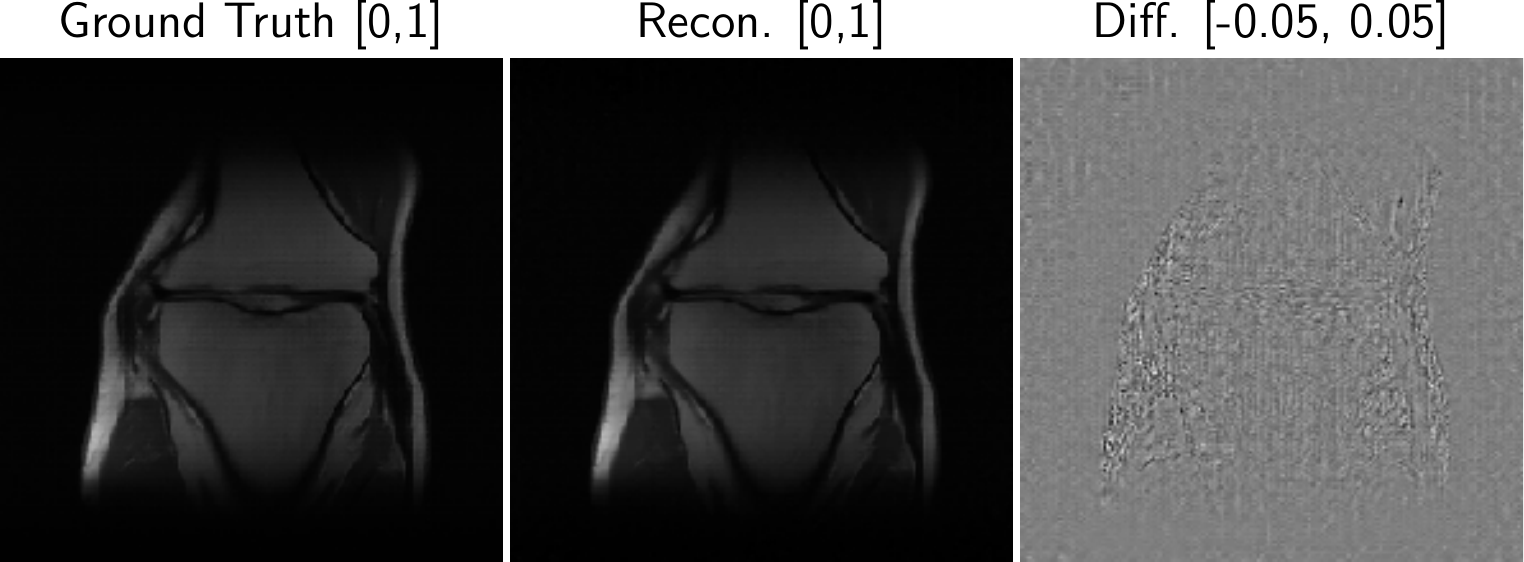}}
\caption{Ground truth, reconstruction and difference plot for reconstruction of a latent-projected image, reconstructed from noiseless, 8 fold undersampled measurements}
\label{fig:recon_trunc}
\end{figure}
\begin{table*}[!t]
    \centering
    \caption{Comparison of RMSE and SSIM for different algorithms for reconstruction from noiseless undersampled data for the simulation study. The metrics are evaluated on a single test image shown in Figures \ref{fig:knee_8x_nn}, \ref{fig:knee_20x_nn}, \ref{fig:brain_8x_nn} and \ref{fig:brain_20x_nn}. The metrics in brackets are evaluated on a region of interest (ROI) selected as shown in Figures \ref{fig:knee_8x_nn}, \ref{fig:knee_20x_nn}, \ref{fig:brain_8x_nn} and \ref{fig:brain_20x_nn}. }
    \label{tab:mse_ssim_noiseless}
    \begin{tabular}{cccccccccccc}
    \toprule
                 & {} & \multicolumn{2}{l}{Knee (in dist.) 8x} & \multicolumn{2}{l}{Knee (in dist.) 20x} & \multicolumn{2}{l}{Brain (in dist.) 8x} & \multicolumn{2}{l}{Brain (in dist.) 20x} \\
                 & {} &          RMSE full &      SSIM full &           RMSE full &      SSIM full &           RMSE full &      SSIM full &            RMSE full &      SSIM full \\
                 & {} &          (ROI) &      (ROI) &           (ROI) &      (ROI) &           (ROI) &      (ROI) &            (ROI) &      (ROI) \\
    \midrule
    \multirow{2}{*}{PLS-TV} & \multirow{8}{*}{{}} &             0.0114 &         0.9793 &              0.0216 &          0.949 &              0.0195 &         0.9672 &               0.0547 &         0.8421 \\
                 & \multirow{7}{*}{{}} &            (0.021) &       (0.9577) &             (0.049) &       (0.8516) &            (0.0232) &       (0.9539) &             (0.0631) &       (0.7689) \\
    \cline{1-10}
    \multirow{2}{*}{CSGM-GAN} & \multirow{6}{*}{{}} &             0.0298 &         0.9266 &              0.0327 &         0.9165 &              0.0545 &         0.8342 &               0.0553 &         0.8293 \\
                 & \multirow{5}{*}{{}} &           (0.0699) &       (0.7589) &            (0.0738) &       (0.7446) &            (0.0733) &       (0.7381) &             (0.0723) &       (0.7543) \\
    \cline{1-10}
    \multirow{2}{*}{IAGAN-TV} & \multirow{4}{*}{{}} &        \bf{0.0101} &    \bf{0.9833} &         \bf{0.0139} &    \bf{0.9727} &         \bf{0.0114} &    \bf{0.9836} &          \bf{0.0238} &    \bf{0.9474} \\
                 & \multirow{3}{*}{{}} &      \bf{(0.0185)} &  \bf{(0.9645)} &       \bf{(0.0285)} &  \bf{(0.9289)} &       \bf{(0.0138)} &  \bf{(0.9785)} &        \bf{(0.0293)} &  (0.9267) \\
    \cline{1-10}
    \multirow{2}{*}{INN Proj. TV} & \multirow{2}{*}{{}} &             0.0113 &         0.9789 &              0.0156 &         0.9666 &               0.014 &         0.9757 &               0.0256 &         0.9398 \\
                 & {} &           (0.0206) &       (0.9589) &            (0.0325) &       (0.9157) &            (0.0165) &       (0.9708) &             \bf{(0.0293)} &       \bf{(0.9289)} \\
    \bottomrule
    \end{tabular}
    
    \end{table*}

\section{FISTA for solving the PLS-TV optimization problem}
The PLS-TV optimization problem -- formulated as
\begin{align}
    \hat{\f} &= \argmin_{\f} \norm{\g - H\f}_2^2 + \lambda\norm{\f}_{\rm{TV}},
\end{align}
where 
\begin{align}
    \norm{\f}_{\rm{TV}} &= \sum_{i,j} (|\f_{i, j} - \f_{i, j+1}| + |\f_{i,j} - \f_{i+1,j}|)
\end{align}
and $\f_{i,j}$ represents the $(i,j)$-th pixel of the image $\f$ -- was solved using FISTA \cite{fista}. The basic step in FISTA, given by Eq. (4.1) in the original paper by Beck, \textit{et al.}, is a composition of a gradient update and a proximal update. The gradient update was performed using Python, with automatic gradient computation from Tensorflow. The proximal update was performed using the Python package ProxTV \cite{proxtv}. For the emulated experimental study, where $\f$ is complex valued, the real and imaginary parts were considered separately for the computation of the proximal update.

\section{Reconstruction from noiseless, fully sampled $k$-space measurements}\label{sec:inv_crime}
This study was conducted to analyze the loss decay during the iterative optimization, when the forward operator $H$ is bijective. Let $\g = H\tilde{\f}$ be the measurement corresponding to the unknown true object $\tilde{\f}$. Figure \ref{fig:fullsamp_recon} displays the results of this study. Figure \ref{fig:inverse_crime} shows the normalized mean square loss versus the iteration. As can be seen in the figure, the loss does not go down to zero to machine precision, but the mean square reconstruction error goes down to around $10^{-6}$.
Thus, similar to other deep learning based methods for image reconstruction \cite{sidky_pan_cnn}, we obtain only an approximate solution to
\begin{align*}\label{eqn:csgm_inn}
    \hat{\z} =& \argmin_{\z} \norm{\g - HG_{\rm{inn}}(\z) }_2^2 - \lambda\log p_Z(\z) \\
    &\qquad\qquad\qquad\qquad\qquad\qquad\quad + \mu\norm{G_{\rm{inn}}(\z)}_{\rm{TV}},\\
    \text{subject to} &\quad \z_{1:n-k} = 0,\\
    \hat{\f} \equiv&~ G_{\rm{inn}}(\hat{\z}),\numberthis{}
\end{align*}
even with complete measurements, which can be attributed to the fact that a non-convex optimization problem is solved by use of gradient-based methods.

\begin{figure*}[!t]
\centerline{\includegraphics[width=0.8\linewidth]{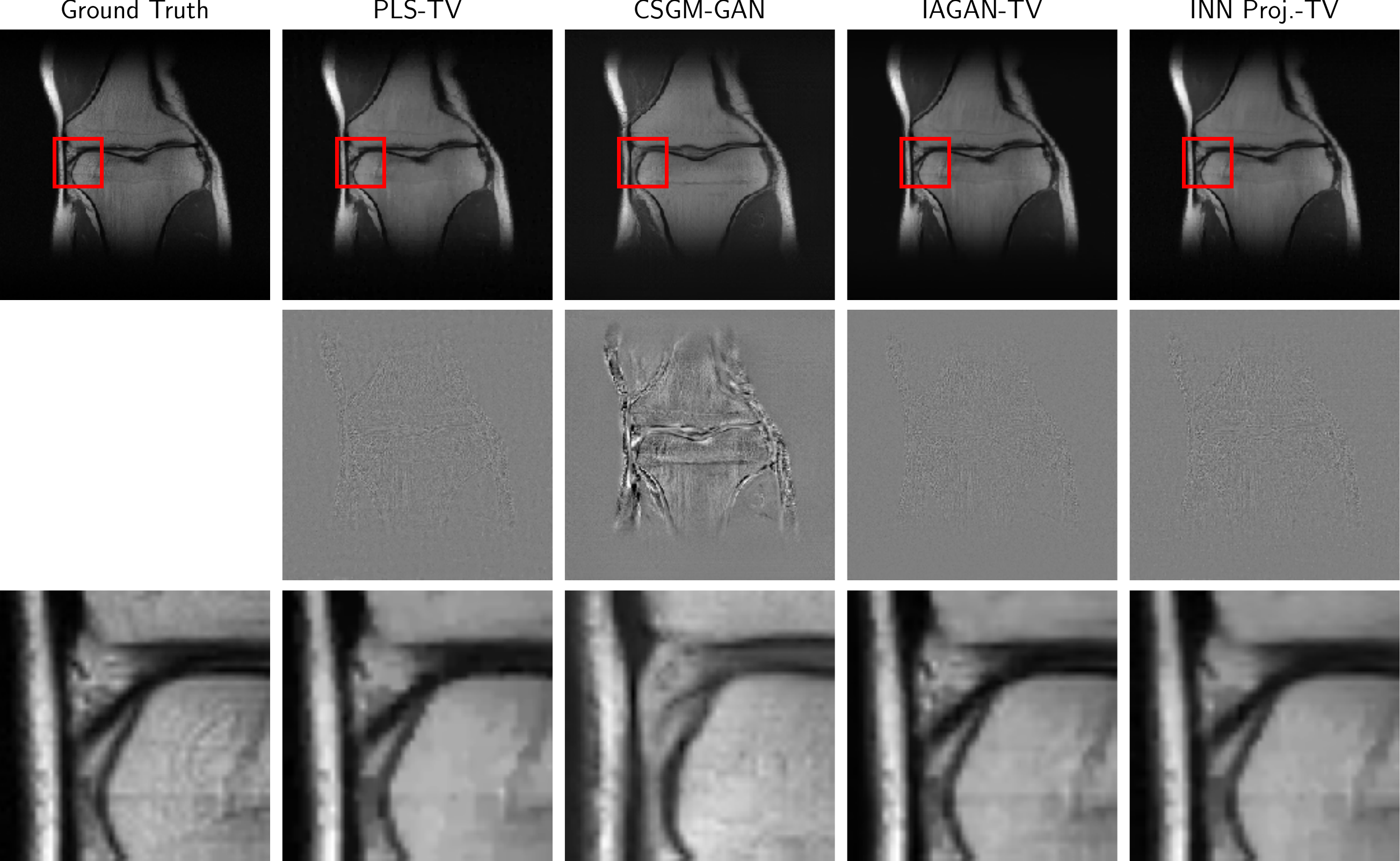}}
\caption{Ground truth, difference plots and reconstruction results for a coronal PD weighted knee image without fat suppression from noiseless 8 fold undersampled measurements.}
\label{fig:knee_8x_nn}
\end{figure*}
\begin{figure*}[!t]
\centerline{\includegraphics[width=0.8\linewidth]{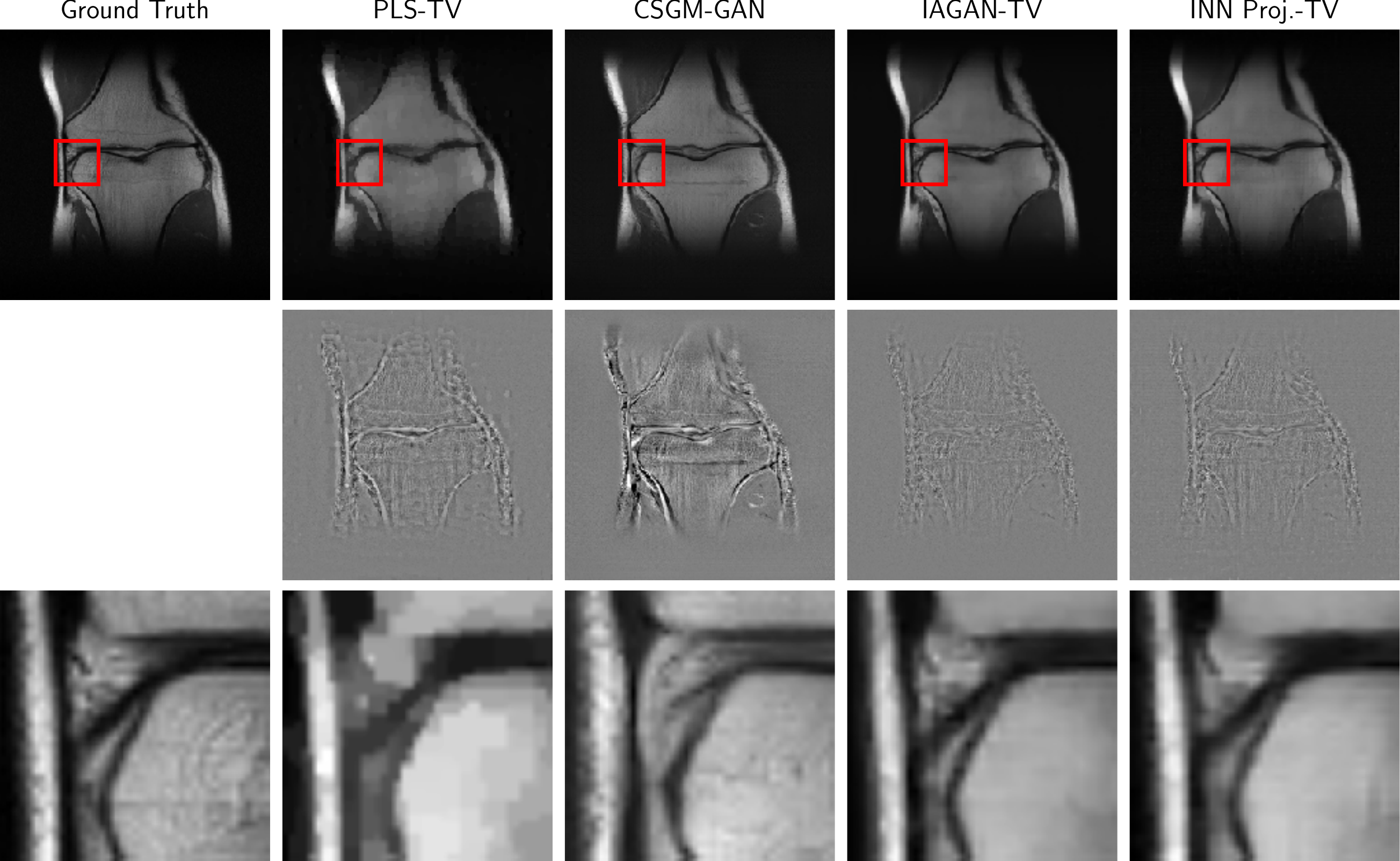}}
\caption{Ground truth, difference plots and reconstruction results for a coronal PD weighted knee image without fat suppression from noiseless 20 fold undersampled measurements.}
\label{fig:knee_20x_nn}
\end{figure*}
\begin{figure*}[!t]
\centerline{\includegraphics[width=0.8\linewidth]{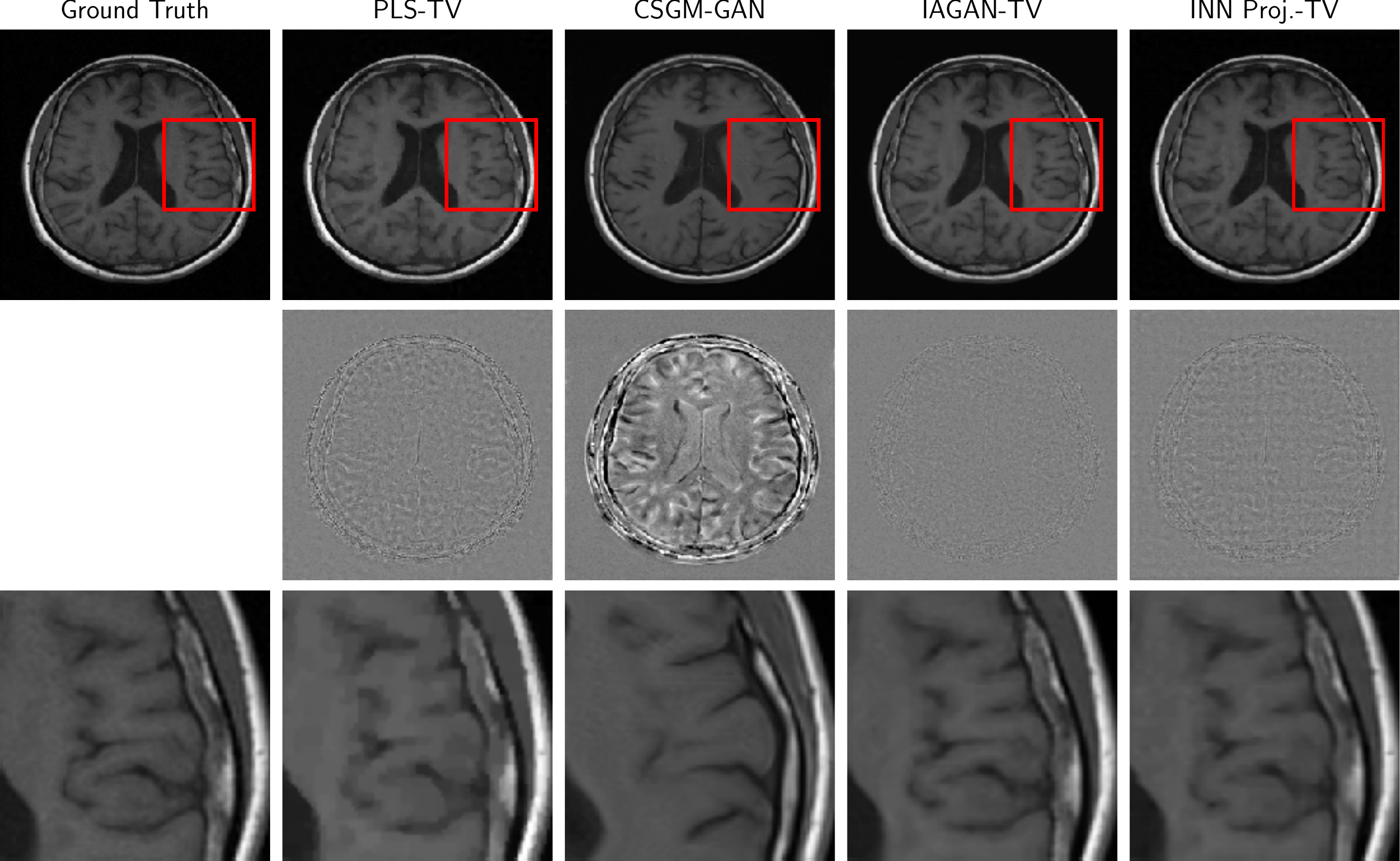}}
\caption{Ground truth, difference plots and reconstruction results for a T1 weighted axial brain image from noiseless 8 fold undersampled measurements.}
\label{fig:brain_8x_nn}
\end{figure*}
\begin{figure*}[!t]
\centerline{\includegraphics[width=0.8\linewidth]{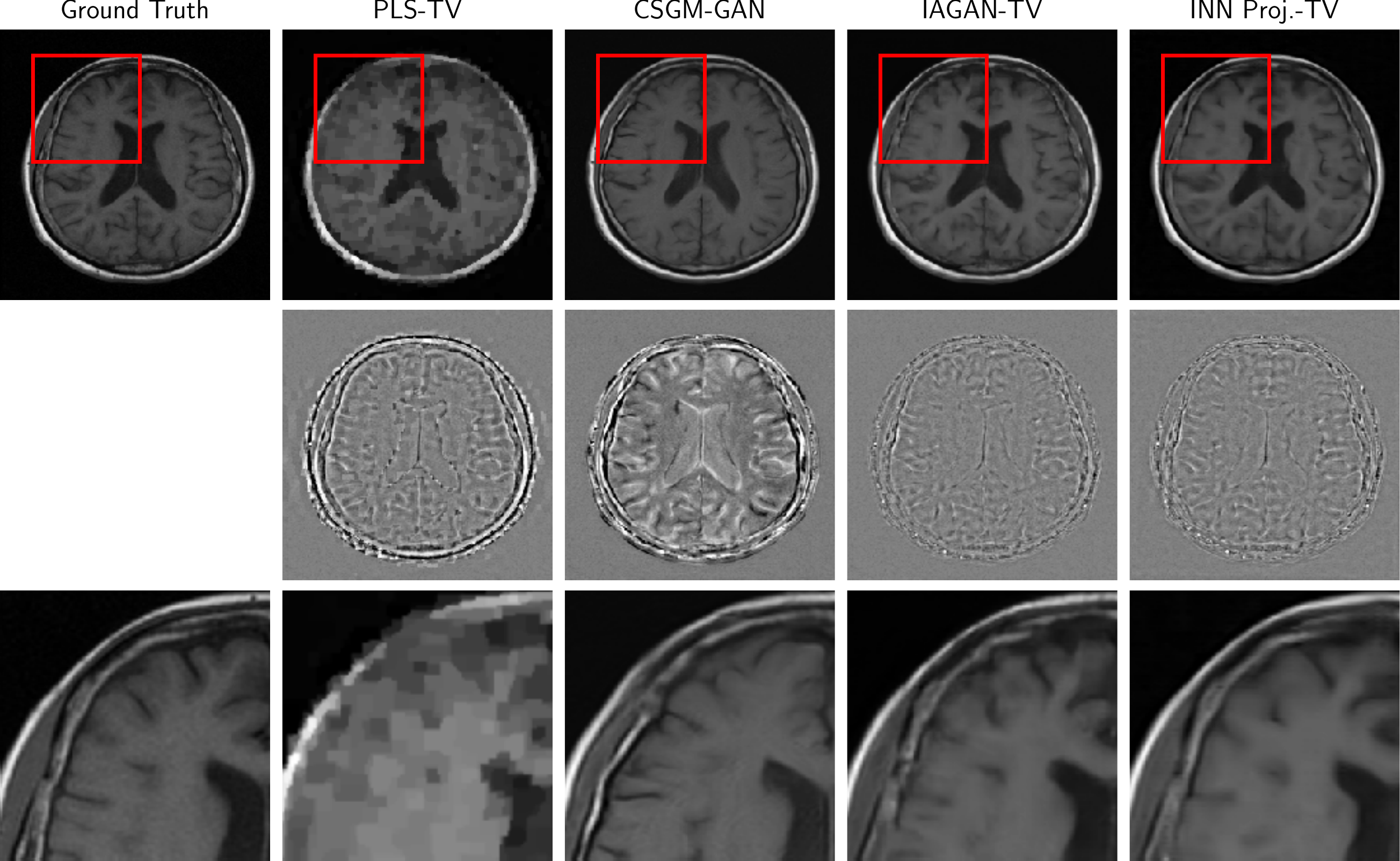}}
\caption{Ground truth, difference plots and reconstruction results for a T1 weighted axial brain image from noiseless 20 fold undersampled measurements.}
\label{fig:brain_20x_nn}
\end{figure*}

\begin{figure}[!t]
\centerline{\includegraphics[width=0.6\linewidth]{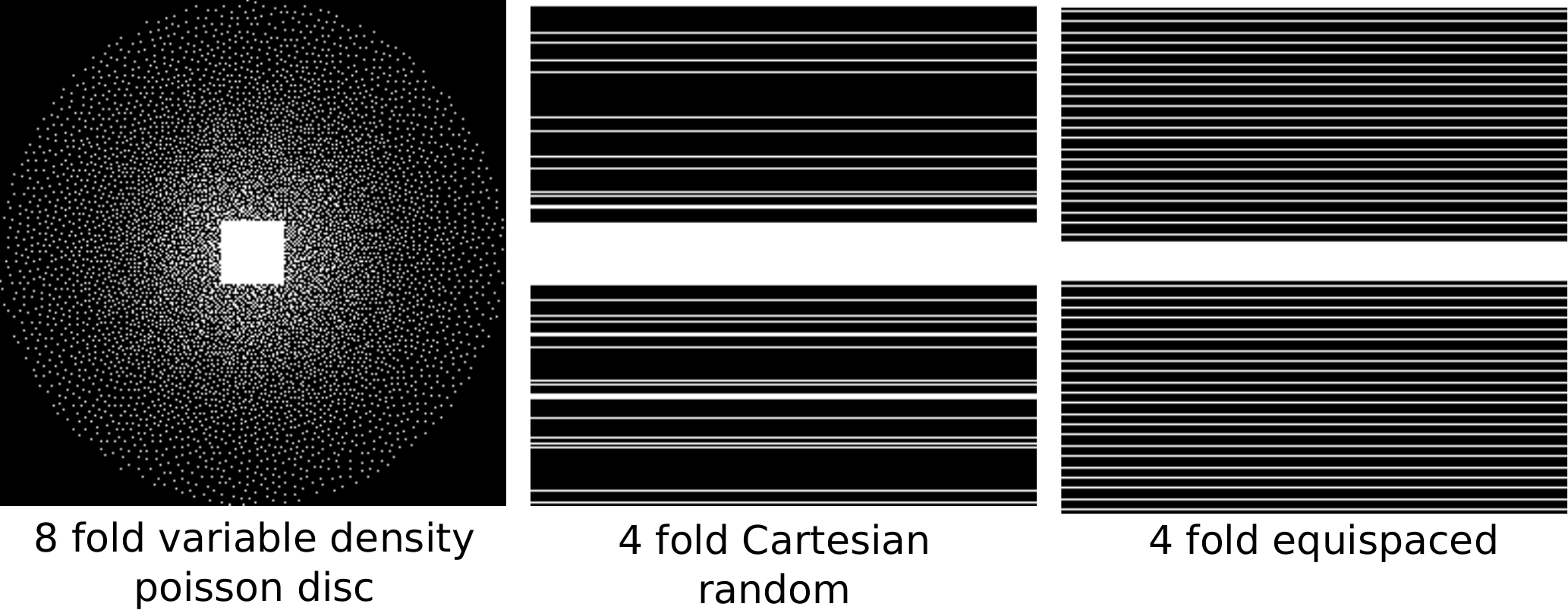}}
\caption{Undersampling masks used for the mask variation study.}
\label{fig:undersampling_masks_3}
\end{figure}

\noindent\begin{figure}
\begin{subfigure}[h]{0.5\linewidth}
\includegraphics[width=\linewidth]{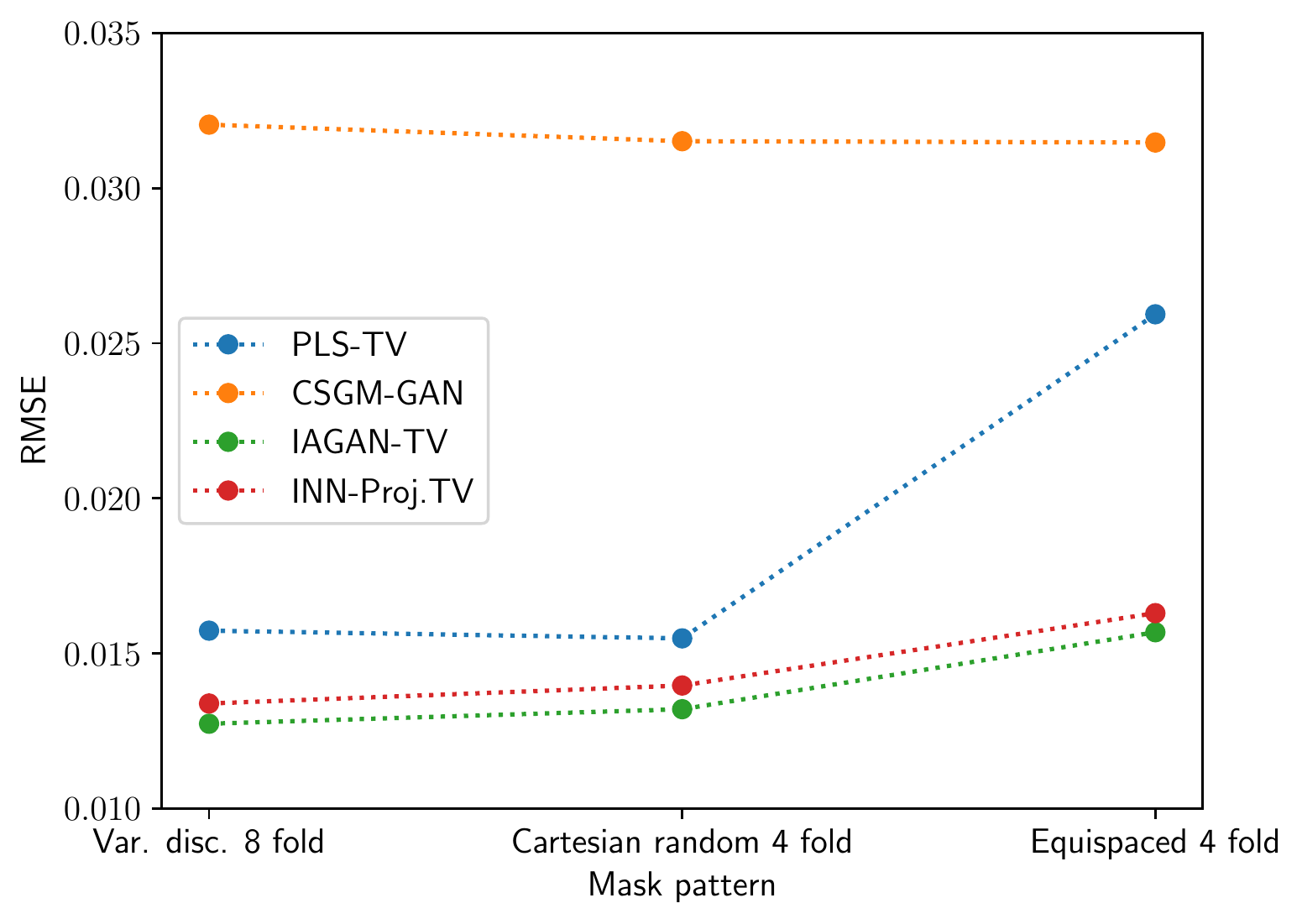}
\captionof{figure}{RMSE performance of the algorithms for different undersampling masks}
\label{fig:mask_variation_rmses}
\end{subfigure}
\begin{subfigure}[h]{0.5\linewidth}
\includegraphics[width=\linewidth]{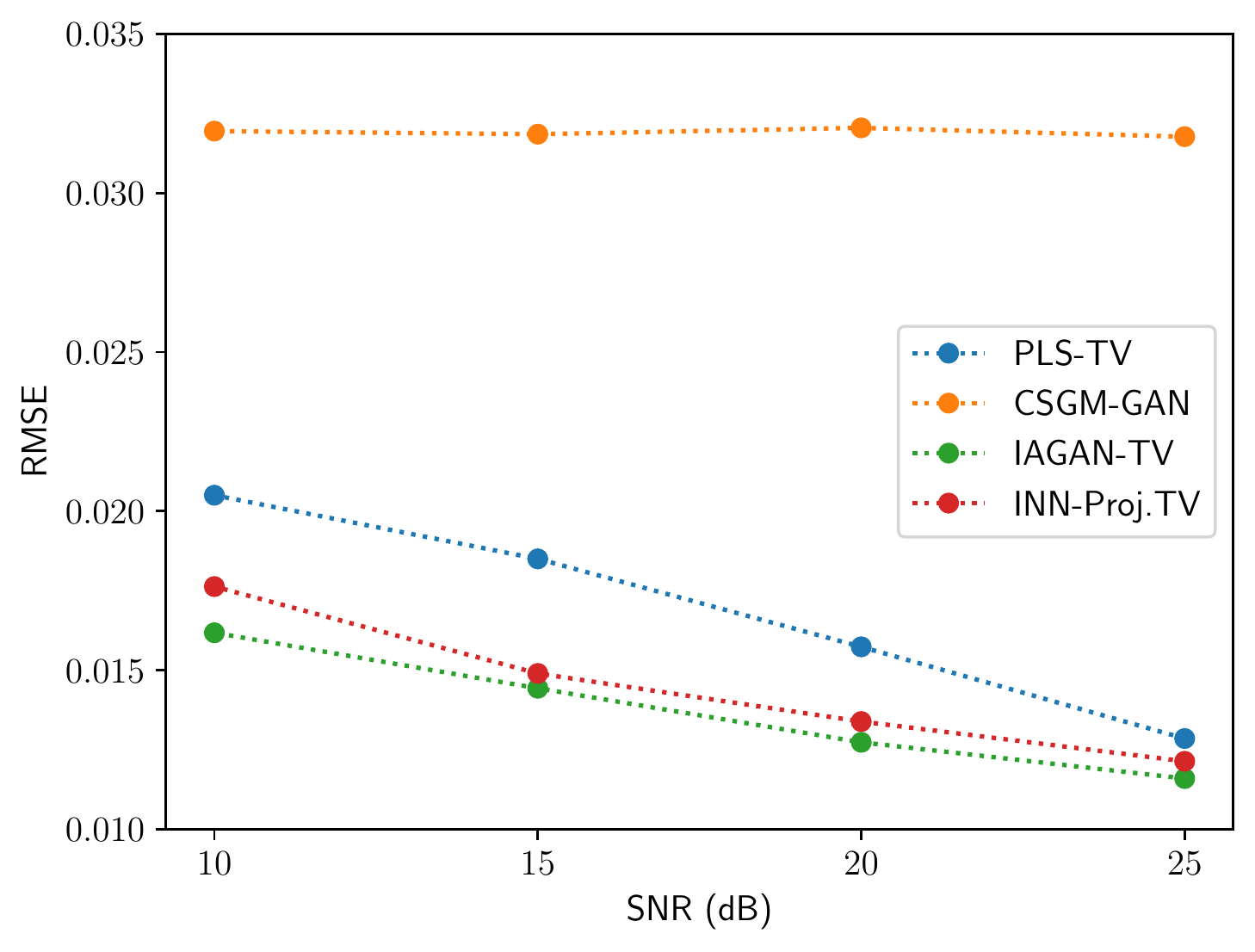}
\captionof{figure}{RMSE performance of the algorithms for different measurement noise levels}
\label{fig:noise_variation_rmses}
\end{subfigure}
\caption{RMSE performance of the algorithms for different undersampling masks and different measurement noise levels.}
\end{figure}

\noindent\begin{figure}
\begin{subfigure}[h]{0.51\linewidth}
\includegraphics[width=\linewidth]{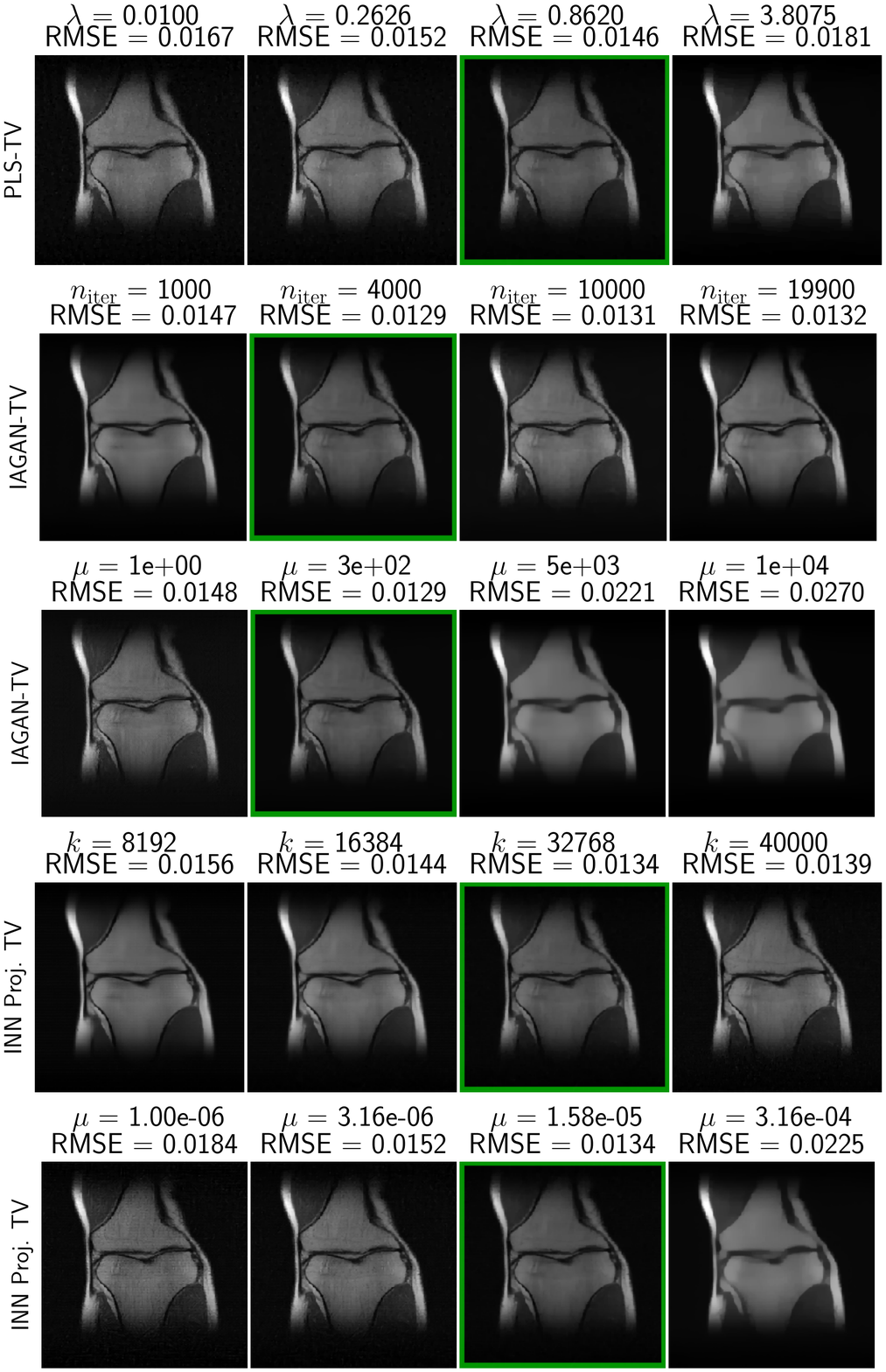}
\captionof{figure}{Reconstruction from 8 fold undersampled measurements}
\label{fig:lamsweep_knee8x}
\end{subfigure}
\begin{subfigure}[h]{0.51\linewidth}
\includegraphics[width=\linewidth]{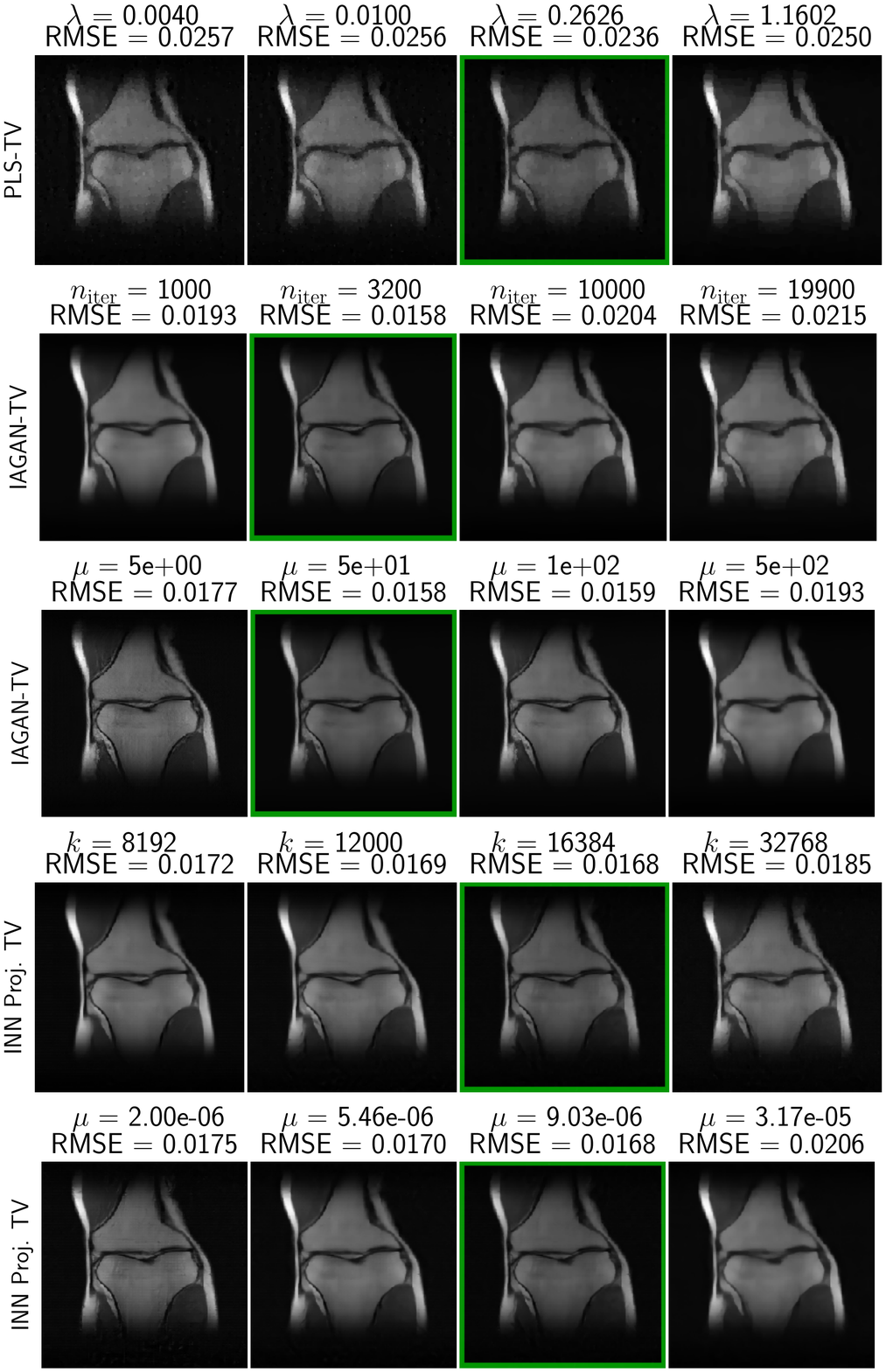}
\captionof{figure}{Reconstruction from 20 fold undersampled measurements}
\label{fig:lamsweep_knee20x}
\end{subfigure}
\caption{{Reconstructed coronal PD weighted knee images from 8 and 20 fold undersampled measurements in the simulation study, for various regularization parameter settings for different reconstruction methods. $\lambda$ denotes the TV regularization weight for PLS-TV, $n_{\rm{iter}}$ denotes the number of iterations for IAGAN-TV (regularization via early stopping), $k$ denotes the dimensionality of the latent subspace for INN Proj. TV, and $\mu$ is used to denote the TV regularization weight for IAGAN-TV and INN Proj. TV.}}
\end{figure}

\noindent\begin{figure}
\begin{subfigure}[h]{0.5\linewidth}
\includegraphics[width=\linewidth]{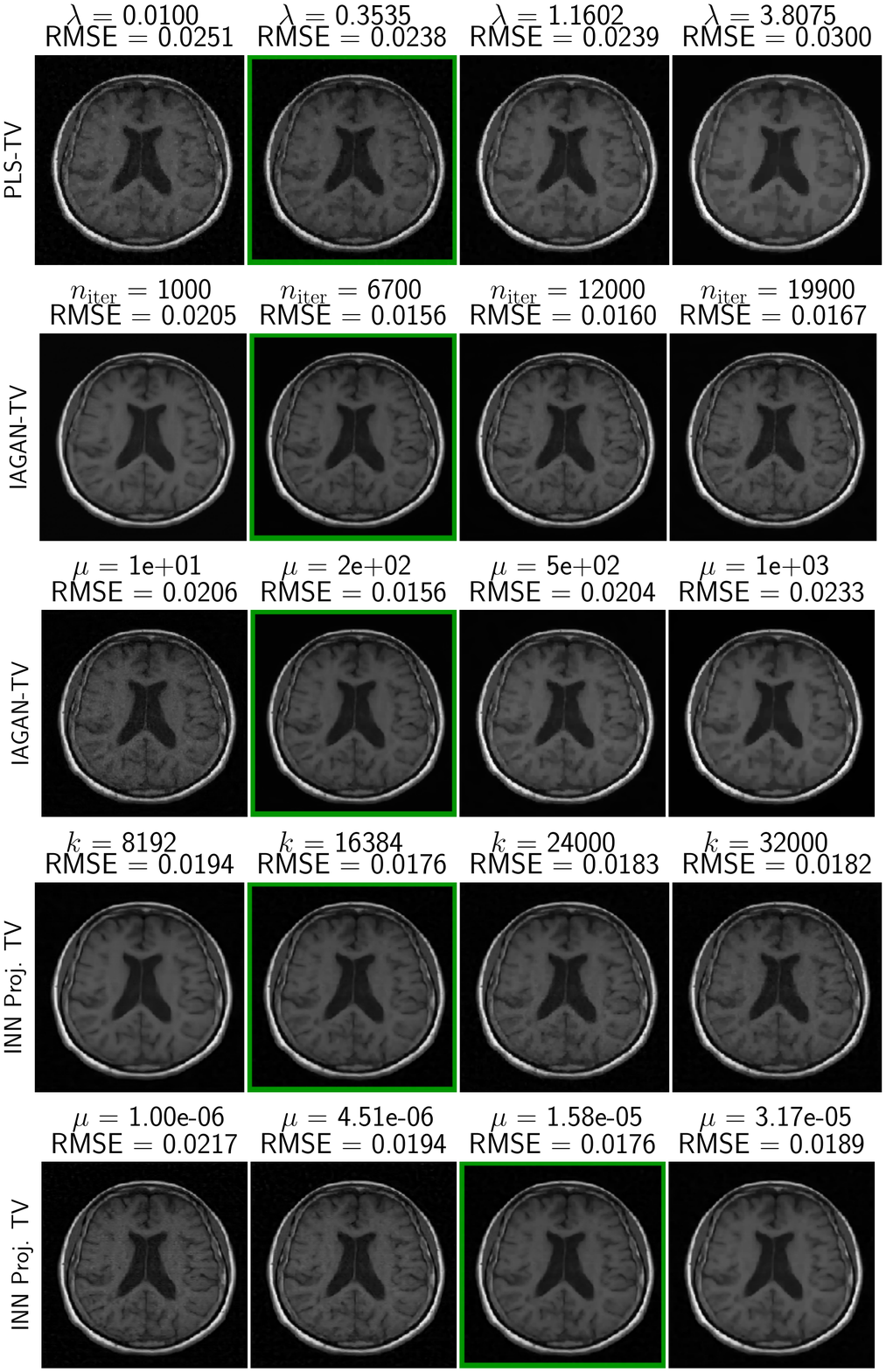}
\captionof{figure}{{Reconstruction from 20 fold undersampled measurements}}
\label{fig:lamsweep_brain8x}
\end{subfigure}
\begin{subfigure}[h]{0.5\linewidth}
\includegraphics[width=\linewidth]{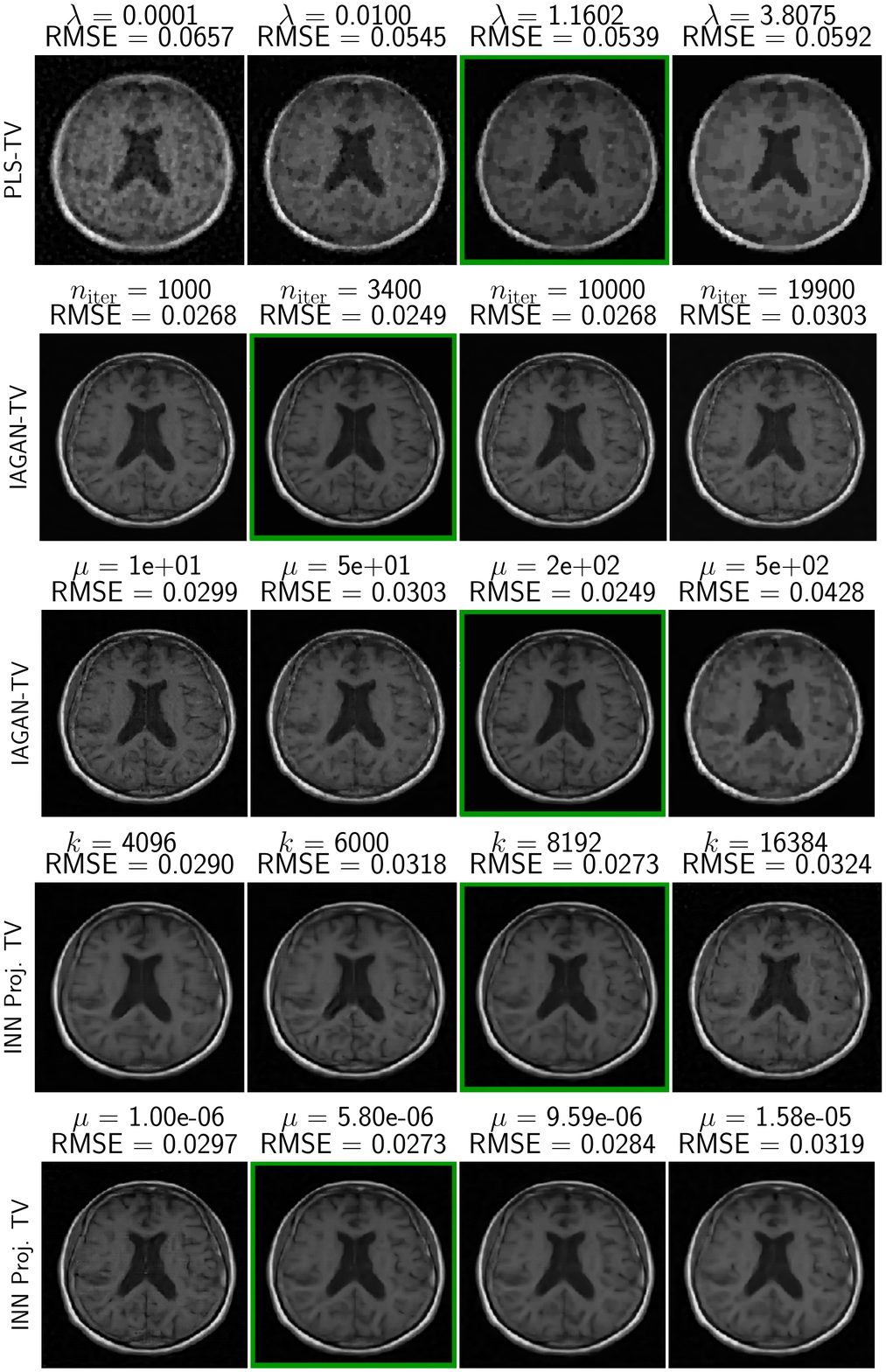}
\captionof{figure}{{Reconstruction from 20 fold undersampled measurements}}
\label{fig:lamsweep_brain20x}
\end{subfigure}
\caption{{Reconstructed axial T1 weighted brain images from 8 and 20 fold undersampled measurements in the simulation study, for various regularization parameter settings for different reconstruction methods. $\lambda$ denotes the TV regularization weight for PLS-TV, $n_{\rm{iter}}$ denotes the number of iterations for IAGAN-TV (regularization via early stopping), $k$ denotes the dimensionality of the latent subspace for INN Proj. TV, and $\mu$ is used to denote the TV regularization weight for IAGAN-TV and INN Proj. TV.}}
\end{figure}
\section{Reconstruction of latent-projected images from noiseless, simulated undersampled measurements}
Here, a ground-truth image $\tilde{\f} \in S_k \cap T_\nu$ is referred to as a latent-projected image, where $S_k = \{ \f ~ s.t. ~ \norm{G_{\rm{inn}}^{-1}(\f)} \leq k \}$, and $T_\nu = \{ \f ~ s.t. ~ \norm{\f}_{\rm{TV}} \leq \nu\}$. Since the measurements are noiseless, this image perfectly satisfies the measurement model. Image reconstruction on an ensemble of 20 images. One of the reconstructed images is shown in \autoref{fig:recon_trunc}.

\section{Simulation study: reconstruction of images from undersampled measurements}
Figures \ref{fig:knee_8x_nn} and \ref{fig:knee_20x_nn} display reconstructed images of a coronal knee image from \textit{noiseless} 8 fold and 20 fold undersampled measurements respectively. Figures \ref{fig:brain_8x_nn} and \ref{fig:brain_20x_nn} display reconstructed images of an axial brain image from noiseless 8 fold and 20 fold undersampled measurements respectively. The corresponding RMSE and SSIM values are shown in \autoref{tab:mse_ssim_noiseless}.

\subsection{Convergence Analysis}
For the reconstruction of a coronal knee image from the 8 fold undersampled measurements with 20 dB measurement SNR, the total loss at iteration $i$, defined by
\begin{align}
    \ell^{(i)} = \frac{1}{n} \norm{\g - HG_{\rm{inn}}(\z^{(i)})}_2^2 + \mu \norm{G_{\rm{inn}}(\z^{(i)})}_{\rm{TV}}
\end{align}
was plotted versus the iteration $i$, in \autoref{fig:loss_vs_iter}.

\section{Mask variation study}
The manuscript describes the performance of the PLS-TV, CSGM-GAN, IAGAN-TV and INN Proj.TV methods for reconstruction of images from measurements generated by use of R = 8 and R = 20 variable density Poisson disc undersampling masks for the simulation study, and 4 fold Cartesian random undersampling mask for the emulated experimental study. Here, an estimate of a single knee image is obtained from simulated stylized MRI measurements that are undersampled using different masks. The masks used for this study are shown in \autoref{fig:undersampling_masks_3}. The corresponding RMSE values for the various reconstruction methods are shown in \autoref{fig:mask_variation_rmses}.

\section{Noise variation study}
Here, an estimate of a single knee image is obtained from simulated stylized MRI measurements that are undersampled using the 8 fold variable density poisson disc undersampling mask, with varying levels of noise added to the measurements. The RMSE performance of the four reconstruction methods for measurement SNRs ranging from 25 dB to 10 dB is shown in \autoref{fig:noise_variation_rmses}.

\section{Regularization parameter selection}\label{sec:reg_sweep}
The regularization parameter $\lambda$ for PLS-TV was chosen by line search, and the parameter giving the best RMSE value was chosen. 
The regularization parameters for IAGAN-TV (i.e. the stopping iteration $n_{\rm iter}$ and the TV regularization weight $\mu$), and those for the proposed method (i.e. the latent subspace dimension $k$ and the TV regularization weight $\mu$) were chosen with the help of a 2D grid search, and the parameters giving the best RMSE value was chosen.
Figures \ref{fig:lamsweep_knee8x}, \ref{fig:lamsweep_knee20x}, \ref{fig:lamsweep_brain8x} and \ref{fig:lamsweep_brain20x} show the selection of the regularization parameters.
From the 2D grid search for IAGAN-TV and the proposed method, only the 1D line-search images through the optimal parameter point are shown in Figures \ref{fig:lamsweep_knee8x}, \ref{fig:lamsweep_knee20x}, \ref{fig:lamsweep_brain8x} and \ref{fig:lamsweep_brain20x}.

{
\bibliographystyle{IEEEtran}
\bibliography{csmri_suppl}
}

\ifCLASSOPTIONcaptionsoff
  \newpage
\fi